%% file: pap2-arxiv.tex
\documentclass[sigconf]{acmart}

	\AtBeginDocument{%
		}
	
	\setcopyright{acmlicensed}
	\copyrightyear{2026}
	\acmYear{2026}
	\acmDOI{XXXXXXX.XXXXXXX}
	\acmISBN{978-1-4503-XXXX-X/2018/06}

	\usepackage{myMac-issac}	
	\usepackage{multirow}	
	\input{lmac}

\renewcommand{\issacArxiv}[2][]{#2}		
\renewcommand{\ignoreInArxiv}[1]{}		
	\usepackage{float}
	\usepackage[ruled]{algorithm2e}
\begin{document}
\title{	Covers for Initial Value Problem:\\
	Complete Validated Algorithms with Complexity Analysis
	}

\author{Bingwei Zhang}
	\authornote{Both authors contributed equally to this research.}
	\affiliation{%
	\institution{The Courant Institute of Mathematical Sciences}
	\city{New York}
	\country{USA}
	}\email{bz2517@nyu.edu}

\author{Chee Yap}
  \thanks{%
  The results of this research was funded by NSF Grant \#CCF-2212462.}
	\authornotemark[1]
	\authornote{Corresponding author}
	\affiliation{%
	\institution{The Courant Institute of Mathematical Sciences}
	\city{New York}
	\country{USA}
	}  \email{yap@cs.nyu.edu}

	\renewcommand{\shortauthors}{Zhang and Yap}
	
	
	

\input{inc/abstract}

\keywords{initial value problem, IVP, end cover problem,
	reachability problem, end enclosure
	problem, validated algorithms, interval methods,
	logarithmic norm, matrix measure, 
}


	\maketitle

    \input{inc/1}

    \input{inc/2}

    \input{inc/3}

    \input{inc/4}
    \input{inc/5}

    \input{inc/6}




\renewcommand{\alt}[2]{#2}
\renewcommand{\alt}[2]{#1} 
\alt{

\input{pap2-arxiv.bbl}
}{
	\bibliographystyle{abbrv}
	\bibliography{st,yap,exact,com,gen,geo,alge,
		math,algo,mesh,tnt,ran,bk/bk,bk/bknew}
}
\input{inc/appendixA}
\input{inc/appendixB}


\end{document}

%% file: lmac.tex
\newcommand{\issacArxiv}[2][]{#1}		
\newcommand{\ignoreInArxiv}[1]{}		

\newcommand{\bwX}[2][]{#1}
\newcommand{\chee}[2][]{\coblue{Chee says: #2}\cocyan{New: #1}}
\newcommand{\cheeX}[2][]{#1}

\newcommand{\End}{\mbox{End}}
\newcommand{\Image}{\mbox{Image}} %
\newcommand{\EndCover}{\mbox{EndCover}} 
\newcommand{\EndCoverAlgo}{\mbox{{\tt EndCover\_Algo}}}
\newcommand{\BoundaryCov}{\mbox{BoundaryCover}} %
\newcommand{\wmax}{w_{\max}} 
\newcommand{\wmin}{w_{\min}} 

\newcommand{\nnorm}[1][\cdot]%
	{\left\lVert #1 \right\rVert_{\max}}

\def\lognorm{\mbox{\rm logNorm}}

\def\ivp{\mbox{{\ttt{IVP}}}}		
\def\stepA{{\mbox{\tt StepA}}}
\def\stepB{{\mbox{\tt StepB}}}

\def\simpleIVPdirect{\mbox{\ensuremath{\tt StepB_{Direct}}}}

\def\endEnc{\mbox{{\tt EndEncl}}} 
\def\endEncIVP{\endEnc} 
\def\endEncAlgo{\mbox{{\tt EndEncl Algorithm}}} 
\def\endEnclAlgo{\mbox{{\tt EndEnclosure Algorithm}}} 
\def\endCovAlgo{\mbox{{\tt EndCover Algorithm}}} %

\def\capdCr{\mbox{\tt CAPD}}

\def\vnodelp{\mbox{\tt VNODE-LP}}

\def\Refine{\mbox{\ttt{Refine}}}

\def\Extend{\mbox{\ttt{Extend}}}

\def\filler{\mbox{\ttt{Filler}}}

\def\stage{{\calS}}

\def\TransformBound{{\ttt{TransformBound}}}
\def\SubrSeven{{\ttt{EulerTube}}}
\def\eulertube{{\ttt{EulerTube}}}
\def\Bisect{{\ttt{Bisect}}}

\def\stage{{\calS}} 

\newcommand{\euler}{^{\text{euler}}}

\newcommand{\valid}{\mbox{\texttt{Valid}}}
\newcommand{\cyl}{\mbox{\texttt{Cy\!l}}}

\DeclareRobustCommand{\looongrightarrow}{%
  \DOTSB \relbar\joinrel \relbar\joinrel
  		 \relbar\joinrel \relbar\joinrel
  		 \relbar\joinrel \relbar\joinrel \rightarrow }

%% file: inc/abstract.tex
\begin{abstract}
	We consider the first order autonomous differential equation
	(ODE) $\bfx'=\bff(\bfx)$ where $\bff: \RR^n\to\RR^n$ is locally
	Lipschitz.
	For a box $B_0\ib\RR^n$ and $h>0$, the set of solutions 
	$\bfx:[0,h]\to\RR^n$ that satisfies $\bfx'(t)=\bff(\bfx(t))$
	and $\bfx(0)\in B_0$ is denoted $\ivp_\bff(B_0,h)$.
	We provide a complete validated algorithm
	for the following 
	\dt{End Cover Problem}: given $(\bff,B_0,\veps,h)$,
	to compute a finite set $\calC$ of boxes such that
		$$End_\bff(B_0,h) ~\ib~ \bigcup_{B\in\calC} B
			~\ib~ \Big(End_\bff(B_0,h) \oplus [-\veps,\veps]^n\Big)$$
	where $End_\bff(B_0,h)=\set{\bfx(h): \bfx\in\ivp(B_0,h)}$.
	We will also give a complexity analysis of our algorithm,
	and
	introduce an alternative technique to compute End Cover $\calC$
	based on covering the boundary of $End_\bff(B_0,h)$.

    We also introduce the \dt{Boundary Cover Problem} to compute a $\veps$-cover for the boundary of $End_\bff(B_0,h)$, leading to a faster algorithm the above End Cover Problem.
	Finally, we give experimental results indicating the
	practicality of our techniques.
\end{abstract}

%% file: inc/1.tex
\sect{Introduction}
\input{inc/1-intro}
\input{inc/1-what-is-new}

\input{inc/1-literature}

\ssect{Paper Overview.}
	The remaining sections of this paper are as follows:
	In Section 2, we first establish some notations and
	review our previous End Enclosure algorithm.
	Then we describe a modification to enhance its performance. 
	This enhanced algorithm  
	is used to construct our End Cover Algorithm.
	In Section 3,
	we introduce the Boundary Cover Problem.
	We prove a key result based on Brouwer's
	Invariance Theorem algorithm that reduces
	it to End Covers in one lower dimension. 
	In Section 4,
	we analyze the complexity of the End Cover Algorithm
	in Sections 2.  
	In Section 5, we describe compare our implementation
	of the End Cover algorithms, comparing them with
	two well-known validated IVP algorithms (CAPD and VNODE).
	We conclude in Section 6 and include an Appendix for proofs.
	\issacArxiv{We show more experiments on Appendix B.}

	{\em The present paper is self-contained. 
	But in \cite{zhang-yap:cover:26arxiv}, 
	we provided more experimental results and other elaborations.
	This pdf file contain hyperlinks for references, theorems, etc.
	}


%% file: inc/1-intro.tex
	The initial value problem (IVP) for ordinary 
	differential equations leads to challenging 
	computational problems. A chief reason is that 
	approximation errors typically grow exponentially 
	large with time. So simulating an IVP over a long 
	time span will break down. Other difficulties include 
	the presence of singularities as well as the 
	phenomenon of stiffness (regions where the solution 
	changes quickly). Although numerical algorithms based 
	on machine precision are very successful in 
	practice, they are often qualitatively wrong.
	So there remains a need for certifiable or 
	\dt{validated algorithms}. Validated IVP algorithms have been 
	studied from the beginning of interval analysis 
	over 50 years ago \cite{moore:bk}. Unfortunately, 
	current validated IVP algorithms are only 
	\dt{partially correct} in the sense that, if they 
	halt, then the output is validated 
	\cite{zhang-yap:ivp:25arxiv}. But halting 
	is rarely addressed. Thus the 
	challenge is to construct \dt{complete validated 
		algorithms}, i.e., halting ones.

	In this paper, we consider the following system of
	first order ordinary differential equations (ODEs)
		\beql{bfx'}
			\bfx' = \bff(\bfx)
		\eeql
	where $\bff=(f_1\dd f_n): \RR^n\to \RR^n$
	is locally Lipschitz.
	For $B_0\ib\RR^n$ and $h>0$, let $\ivp_\bff(B_0,h)$
	denote the set of total differentiable
	functions $\bfx=(x_1\dd x_n):[0,h]\to\RR^n$
	such that $\bfx(0)\in B_0$ and the time derivative
	$\bfx'=(x'_1\dd x'_n)$ satisfies \refeq{bfx'}.
	Call each $\bfx\in \ivp_\bff(B_0,h)$
	a \dt{solution} to the \dt{IVP instance} $(\bff,B_0,h)$
	or\footnote{
		Note that we often omit $\bff$, as it is usually fixed
		or implicit in a given context.
	} $(B_0,h)$.
	The \dt{end slice} and \dt{image} of $\ivp_\bff(B_0,h)$ are 
		\beql{endSet}
			\mmatx{
			\End_\bff(B_0,h) &\as&
				\set{\bfx(h): \bfx\in \ivp_\bff(B_0,h)}\\
			\Image_\bff(B_0,h) &\as&
				\set{\bfx(t): \bfx\in \ivp_\bff(B_0,h), t\in[0,h]}}
		\eeql
	If $B_1$ contains $\End_\bff(B_0,h)$ (resp., $\Image_\bff(B_0,h)$),
	we call $B_1$ an \dt{end enclosure} (resp., \dt{full enclosure})
	of $\ivp_\bff(B_0,h)$.
	In our previous paper
	\cite{zhang-yap:ivp:25arxiv},
	we gave a complete validated algorithm for the
	\dt{End Enclosure Problem}\footnote{
		Computational problems are specified using a header of the form
			$$\mbox{\tt PROBLEM}(\bfa ; \coblue{\bfb})\to \bfc$$
		where $\bfa,\coblue{\bfb},\bfc$ are sequences representing
		(respectively) the required arguments, the optional
		arguments, and the outputs.
		Default values for optional arguments may
		be directly specified (as in $h=1$ above) or they
		may be context dependent.
		In \cite{zhang-yap:ivp:25arxiv}, 
		$\bfp_0$ was implicitly defaulted
		to the midpoint of $B_0$.
	}
	which is defined by this header:

	\renewcommand{\alt}[2]{#2} 
	\renewcommand{\alt}[2]{#1} 
	\alt{
		\beql{endEnc}
			\includegraphics[width=0.92\columnwidth]{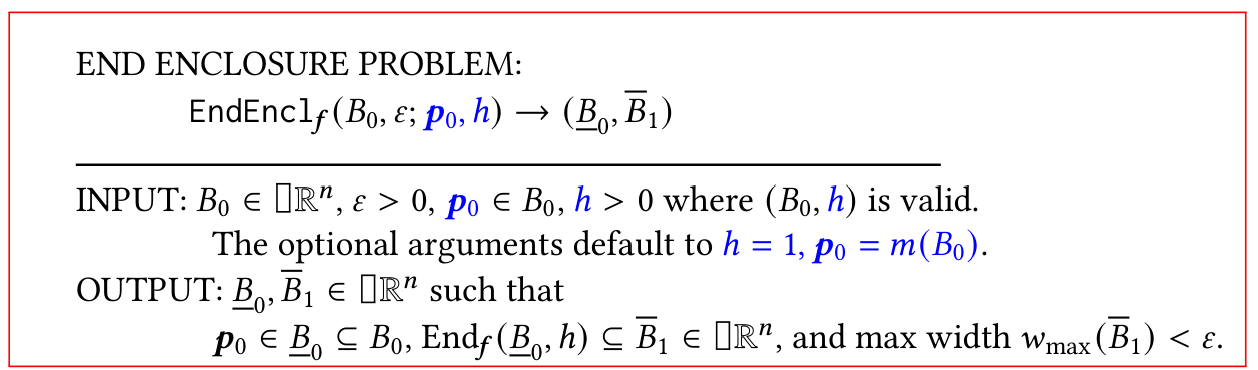}
		\eeql
	}{
	\begin{figure*}
		\Ldent\progb{
			\lline[0]
					\mbox{\rm END ENCLOSURE PROBLEM:}
			\lline[10]
					$\endEncIVP_\bff(B_0,\veps; \coblue{\bfp_0, h})
						\to (\ulB_0,\olB_1)$
			\lline \myhlineY
			\lline[0] INPUT:
				$B_0\in\intbox\RR^n$, $\veps>0$,
					$\coblue{\bfp_0}\in B_0$, $\coblue{h}>0$
				where $(B_0,\coblue{h})$ is valid.
			\lline[12] The optional arguments 
					default to \coblue{$h=1, \bfp_0=m(B_0)$}.
			\lline[0] OUTPUT: $\ulB_0,\olB_1\in\intbox\RR^n$ such that
			\lline[12] $\bfp_0\in \ulB_0 \ib B_0$,
						$\End_\bff(\ulB_0,h) \ib \olB_1\in\intbox\RR^n$,
						and max width $\wmax(\olB_1)<\veps$.
		}
	\end{figure*}
	}

	Here, $\intbox\RR^n$ denotes the set of compact boxes in $\RR^n$,
	and our algorithms assume that the $B$'s in input and output
	are boxes.
	The usual IVP formulation in the literature
	assumes the singleton $B_0=\set{\bfp_0}$.
	Our formulation with a box $B_0$
	(following Corliss \cite[Section 3]{corliss:survey-ode-intvl:89})
	is more realistic since, in real applications
	(e.g.\ biology or physics),
	one never knows an exact point $\bfp_0$,
	only a range of interest.

	A key insight for proving halting of our algorithm
	is to formulate Problem~\refeq{endEnc} as
	a {\em promise problem}:
	the input $(B_0,h)$ is promised to be \dt{valid}.
	This means that, for all $\bfp_0\in B_0$,
	the ODE~\refeq{bfx'} has%
	\footnote{
		Without such a promise, the algorithm would be required
		to output ``invalid'' when no solution exists.
		This would {\em a fortiori} decide validity of the input,
		a decision problem that remains open
		\cite{zhang-yap:ivp:25arxiv}.
	}
	a unique solution $\bfx:[0,h]\to\RR^n$
	with $\bfx(0)=\bfp_0$.

	The output requirement $w(\olB_1)<\veps$
	implies that $B_0$ must be allowed to shrink
	to some $\ulB_0$ (while keeping $\bfp_0\in\ulB_0$).
	In this paper, we address a stronger variant
	in which shrinkage of $B_0$ is not allowed.
	This is the \dt{End Cover Problem}
	defined by this header:

	\renewcommand{\alt}[2]{#2} 
	\renewcommand{\alt}[2]{#1} 
	\alt{
		\beql{endCover}
			\includegraphics[width=0.8\columnwidth]{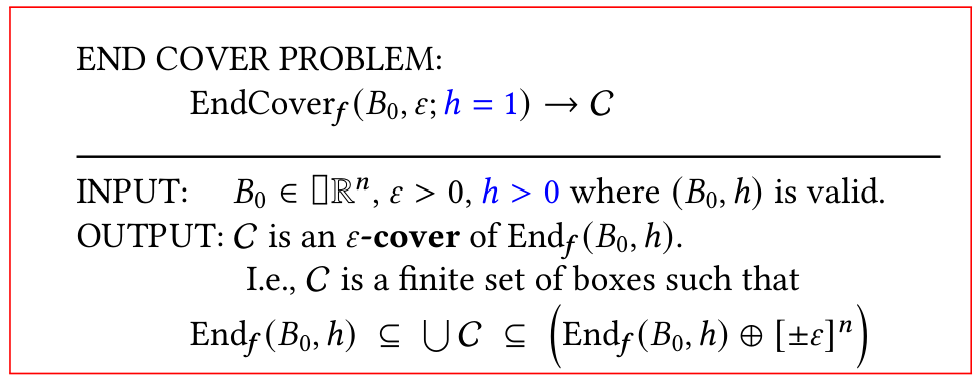}
		\eeql
	}{
		\Ldent\progb{
			\lline[0]
					\mbox{\rm END COVER PROBLEM:}
			\lline[10]
					$\EndCover_\bff(B_0,\veps; \coblue{h=1}) \to \calC$
			\lline \myhlineY
			\lline[0] INPUT: \quad $B_0\in\intbox\RR^n$,
						 $\veps>0$, \coblue{$h>0$}
				where $(B_0,h)$ is valid.
			\lline[0] OUTPUT: $\calC$ is
					an \dt{$\veps$-cover} of $\End_\bff(B_0,h)$.
			\lline[15] 
					I.e., $\calC$ is a finite set of boxes such that
			\lline[10] 
					$\End_\bff(B_0,h) ~\ib~ \bigcup\calC ~\ib~ 
						\Big(\End_\bff(B_0,h) \oplus [\pm \veps]^n\Big) $
		}
	}
	Here, $[\pm \veps]^n$ denotes
	the box $\prod_{i=1}^n [-\veps,\veps]$, 
	$A \oplus B$ is the Minkowski sum of sets $A, B\ib \RR^n$,
	and $\bigcup\calC \as \bigcup_{B\in\calC} B$ is a subset of $\RR^n$.
	For any $\veps>0$, clearly $\veps$-covers of $\End_\bff(B_0,h)$
	exists.  Our introduction of {\em a priori} $\veps$-bounds
	in the problems \refeq{endEnc} and \refeq{endCover}
	is clearly highly desirable.  As far as we know, current IVP
	algorithms do not offer such $\veps$ guarantees. Our use of $\veps$-covers overcomes a difficulty that was pointed out in Eijgenraam's thesis \cite[Example 6.4.2, page 127]{eijgenraam:ivp:81} (and Nedialkov's thesis \cite[p.~77]{nedialkov:thesis:99}.  We reproduce this example in Appendix C.

%% file: inc/1-what-is-new.tex
\ssect{What is new in this paper}
	We have three main objectives:
	The first is to enhance our end enclosure algorithm in
	\cite{zhang-yap:ivp:25arxiv},
	leading to a solution for the End Cover Problem \refeq{endCover}.
	The issue is this: suppose
	we want to compute an $\veps$-cover
	for the IVP instance $(B_0,\veps,h)$ where $B_0=[a,b]$ ($n=1$).
	Using our End Enclosure Algorithm,
	we can compute $B_1$ that is an $\veps$-cover
	of $\End_\bff([a,a_1],h)$ for some $a<a_1\le b$. 
	If $a_1<b$, then we repeat this process
	to compute a $B_2$ to cover $\End_\bff([a_1,a_2],h)$ for
	some $a_1<a_2\le b$, etc.
	If this process terminates, we have a finite 
	$\veps$-cover $\calC=\set{B_1,B_2,\ldots}$.
	Viewing $\bfx$ as space variables, this shows that
	our new problem amounts to computing a ``space cover'' of $[a,b]$,
	analogous to the ``time cover'' of $[0,h]$ in the original
	End Enclosure Problem.

	\begin{figure}
		\centering
		\includegraphics[width=0.7\linewidth]
			{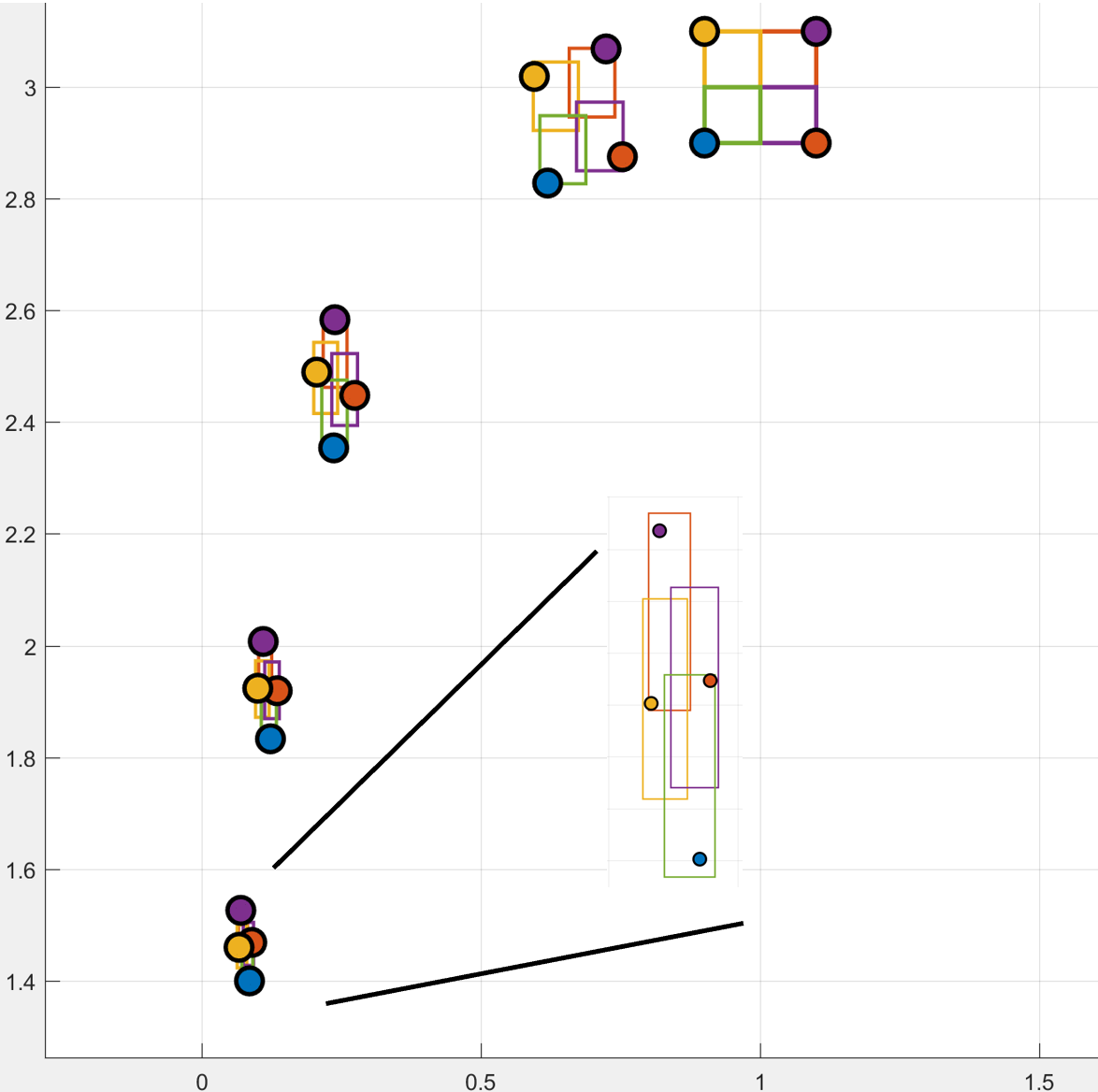}
		\caption{$\veps$-End Covers ($\veps=0.1$) at times
			$h_i\in$ $(0, 0.1, 0.4, 0.7,1.0)$
			for $i=0\dd 4$.  The boxes in $\calC$
			are colored for visualization.
			An enlarged image for the cover at $h_4 = 1.0$ is also shown.
			}
		\label{fig:eg0-eps-endCover}
	\end{figure}
	In \refFig{eg0-eps-endCover}, we show the output of our end cover
	algorithm on the Volterra-Lotka (or predator-prey) system
	(\refTab{problems}, Section 5) with $\veps=0.1$.
	We show five IVP instances ($i=0\dd 4$)
		$$(\bff,B_0,h_i,\veps)
			= \Big(\mmat{2x(1-y)\\-y(1-x)},
				Box_{(1,3)}(0.1),
				h_i,
				0.1	\Big)
			$$
	where $h_i\in (0, 0.1, 0.4, 0.7,1.0)$ and
	$Box_{\bfp}(w)$ denotes box $\bfp+ [-w,w]^2$.

	\cheeX{23Jan: redo the instance of Fig1 using the End Cover
	Algo}

	Our second objective is to introduce an alternative 
	approach to computing end covers:
	the idea is to reduce the covering of
	$\End_\bff(B_0,h)$ to the covering of
	its boundary, $\partial(\End_\bff(B_0,h))$.
	Any $\veps$-cover of $\partial(\End_\bff(B_0,h))$ is called
	an \dt{$\veps$-boundary cover} of $\End_\bff(B_0,h)$.
	Thus we reduce the
	$n$-dimensional problem to a $(n-1)$-dimensional one,
	modulo a \dt{filler problem}: 
	given $\calC$, an $\veps$-boundary cover of $\End_\bff(B_0,h)$,
	to compute a finite set of boxes $\calC_0$ so that
	$\calC\cup\calC_0$ is an $\veps$-cover of $\End_\bff(B_0,h)$.
	Intuitively, $\calC_0$ covers
	the ``interior'' of $\bigcup \calC$.
	 
\begin{figure}
		\centering
		\includegraphics[width=0.7\linewidth]
			{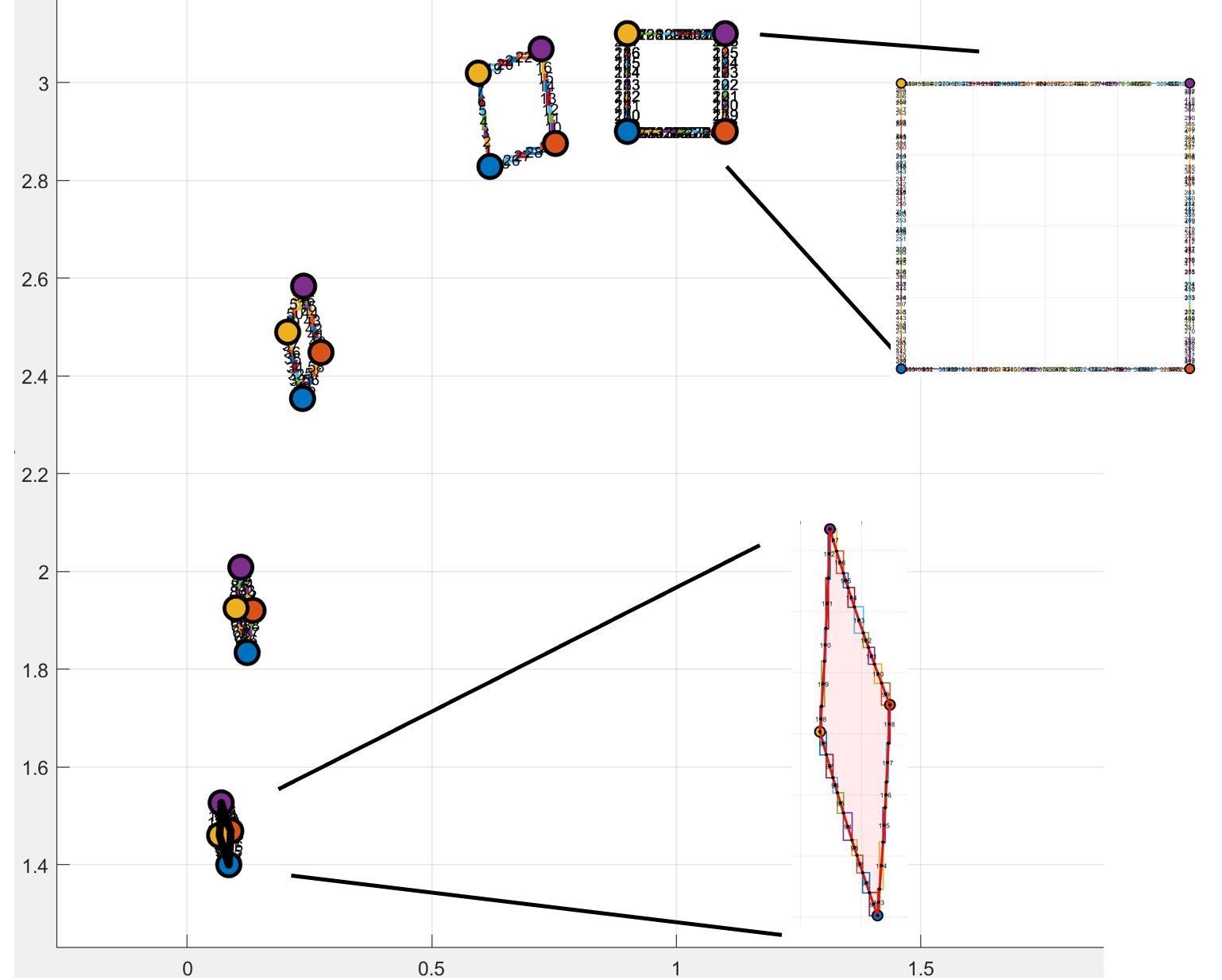}
		\caption{$\veps$-Boundary Covers ($\veps=0.01$)
			at times
			$h_i\in (0, 0.1, 0.4, 0.7,1.0)$
			for $i=0\dd 4$.  The boxes in a cover $\calC$
			are colored to visualize them.
			An enlarged image for the cover at $h_4 = 1.0$ is also shown.
			}
		\label{fig:eg0-eps}
	\end{figure}

	In \refFig{eg0-eps}, we run our boundary cover algorithm
	on the same Volterra-Lotka instances of
	\refFig{eg0-eps-endCover}, but with $\veps=0.01$ instead of
	$\veps=0.1$. 
	The running time is much faster than that of the end cover algorithm.
	The filler problem to convert $\calC_i$ into
	an end cover of $\End_\bff(B_0,h_i)$ is relatively easy
	for these instances because the ``interior polygon''
	(pink area in the enlarged image)
	of $\calC_i$ are simple polygons.
	\issacArxiv{
	We reduced $\veps$ to $0.01$ (from the previous $0.1$)
	to ensure the interior polygons are non-empty.  
	Despite the smaller $\veps$,
	the boundary computation is much faster than the end cover
	computation.
	}
	
	\cheeX{
	We need an example where this boundary method is
	faster!  Bw thinks that if we do longer time simulation,
	we may see it...}
	
	Although the boundary method seems like
	a natural suggestion, its justification
	depends on a non-trivial result\footnote{
		See
	\myHref{https://terrytao.wordpress.com/2011/06/13/brouwers-fixed-point-and-invariance-of-domain-theorems-and-hilberts-fifth-problem/}
	{Terry Tao's blog}
	on Brouwer's Fixed Point Theorem.
	}
	called the
	\dt{Invariance of Domain Theorem} from L.E.J. Brouwer.

	\issacArxiv{
	We use this result to prove 
	\refSec{cover} 
	that if $B_0$ is homeomorphic to a ball
	(e.g., a box), then $\End_\bff(B_0,h)$ is
	also  homeomorphic to a ball. So, if $\calC$ is an $\veps$-boundary
	enclosure of $\End_\bff(B_0,h)$, then $\bigcup\calC$ is
	a connected set which can be ``filled'' by a set 
	$\calC_0$ of boxes to give an $\veps$-end cover, as noted
	above.}

	Our third objective is to give a complexity analysis
	of our end cover algorithm.  Note that
	complexity analysis is rarely attempted in the
	validated algorithms literature for various reasons.
	Using the enhanced End Enclosure algorithm
	(Section 2.3)
	we can now give explicit bounds, based on implicit
	parameters which can be derived from the actual inputs.

%% file: inc/1-literature.tex
\ssectL[literature]{Brief Literature Review and Background.}
	For the validated IVP literature,
	we refer to the surveys by
	Corliss \cite{corliss:survey-ode-intvl:89},
	Nedialkov et al.~\cite{nedialkov+2:validated-ode:99},
	and other references in \cite{zhang-yap:ivp:25arxiv}.
	\ignore{
	These algorithms follow a standard pattern
	based on two subroutines, called
	``Algorithm~I'' and ``Algorithm~II'' by Corliss.
	We denote them by \stepA\ and \stepB,
	emphasizing that they are steps in an
	iterative algorithm.
	Briefly, \stepA\ takes a sufficiently small
	time step so that the ODE remains valid and
	a full enclosure can be estimated,
	while \stepB\ refines this estimate to obtain
	a tighter end enclosure.
	In their complete validated algorithm
	for End Enclosure Problem \cite{zhang-yap:ivp:25arxiv},
	we introduce a scaffold data structure in order to
	to support more flexible algorithmic processes.
	The current paper shows the power of	
	}
	The computational problems associated with $\End(B_0)$
	have many applications, and are often called
	\dt{reachability problems}
	in non-linear control and verification
	(e.g., \cite{shen+2:tight-reach:21,fan+2:fast-controller:20}).
	The notions of partially correct algorithms
	and promise problems are standard in
	theoretical computer science.
	Partial correctness is useful since we can now
	split correctness into separate concerns:
	halting and correct output.
	Promise problems \cite{goldreich:promise-problems:06}
	are also natural in numerical computation,
	where typical promises assert non-singularity
	or good conditioning.
	Algorithms for promise problems may assume
	the promised condition without checking it.
	
	Ultimately, validated algorithms require
	arbitrary-precision numerics
	(e.g., big-number packages such as \ttt{gmp}).
	Although this is our goal, we avoid a direct
	discussion of this aspect in order to focus on
	the main algorithmic development.
	Following standard practice, we describe an
	\dt{abstract algorithm} $A$ that operates on exact
	mathematical objects.  But
	such abstract descriptions must ultimately be
	replaced by implementable operations.
	Following \cite{xu-yap:real-zero-system:19},
	this is done in two steps:
	first, $A$ is transformed into its
	\dt{interval version} $\intbox A$,
	which computes with bounding intervals.
	Since $\intbox A$ is still abstract (e.g.,
	the interval bounds may involve real numbers), we need
	to convert $\intbox A$
	to an \dt{effective version} $\wtbox A$,
	that uses only implementable operations
	(e.g., bigFloats) and finite representations.
	Once $A$ is given (as in this paper), the
	transformations $A \to \intbox A \to \wtbox A$
	are usually straightforward, as global properties
	are largely preserved.
	This Abstract/Interval/Effective approach is outlined in
	\cite{xu-yap:real-zero-system:19}.
	However, not every abstract algorithm $A$
	admits a simple interval version $\intbox A$.
	The difficulty arises when $A$ relies on a ``zero problem''
	(see \cite{yap-sharma:encyclopedia:08}).
	For example, Gaussian elimination for computing
	determinants relies on knowing that a pivot
	is non-zero, and so no simple $\intbox A$ exists.
	Our IVP algorithms avoid such zero problems.

\issacArxiv{
	Our notion of trajectory bundles in \refSSec{traj} appears
	to be new; but the term has appeared in 
	contexts such as in path planning, optimization and optics.
	But these refer to the algorithmic technique of
	 ``multiple shooting or trajectory sampling''
	to better estimate the end set.
	E.g., 
	\cite{bachrach:traj-bundle:thesis13,tracy+6:traj-bundle:25}.}

%% file: inc/2.tex
\sect{The End Cover Problem} 
This section will describe an enhancement of the \endEncAlgo\ in 
\cite{zhang-yap:ivp:25arxiv}. 
We then use the enhanced version
as a subroutine to solve End Cover Problem.
The enhancement is also needed for its complexity analysis.
But first, we establish some notations.

\input{inc/2-notations}

\input{inc/2-algoReview}

\input{inc/2-enhancedEE}
\input{inc/2-endCoverAlgo}


%% file: inc/2-notations.tex
\ssectL[notation]{Notations and Basic Concepts}
	We collect the basic terminology here for easy reference,
	largely following \cite{zhang-yap:ivp:25arxiv}.
	We use bold font like $\bfp=(p_1\dd p_n)$ for vectors,
	with $(\bfp)_i=p_i$ as the $i$th coordinate function.
	The origin in $\RR^n$ is denoted $\0=(0\dd 0)$.
	\ignore{ not used:
		$\bfp\ge \bfq$ means $(\bfp)_i\ge (\bfq)_i$ for $i=1\dd n$.
	Let $\max\bfr=\max_{i=1}^n (\bfr)_i$ and similarly for $\min\bfr$. 
	}
	Let $\intbox\RR$ denote the set of compact intervals.
	We identify the interval $[a,a]$ with $a\in\RR$ and thus
	$\RR\ib\intbox\RR$.
	For $n\ge 1$, let $\intbox\RR^n \as (\intbox\RR)^n$ denote the set 
	of (axes-aligned) boxes in $\RR^n$.
	For a non-empty set $S\ib\RR^n$, let $Ball(S)$ and $Box(S)$
	denote, respectively, the smallest ball 
	and smallest  box containing $S$. Also, let $\bfm(S)$ denote the
	midpoint of $Box(S)$.  For $\alpha\ge 0$, the
	$\alpha$-dilation operator is
	$\alpha S \as \set{\alpha \bfp: \bfp\in S}$.
	E.g., if $n=1$ and $S=[2,6]$ then $\half S =[1,3]$.

	The $Box$ and $Ball$ operators in $\RR^n$
	can also take numerical parameters:
	$Ball(r)$ is the ball in $\RR^n$ centered at $\0$ of radius $r\ge 0$,
	and $Box(\bfr)\as \prod_{i=1}^n[-r_i,r_i]$
	where $\bfr=(r_1\dd r_n)\ge \0$.
	If $\bfp\in\RR^n$, let
		$Ball_\bfp(\bfr)\as \bfp+Ball(\bfr)$ and
		$Box_\bfp(\bfr)\as \bfp+Box(\bfr)$.
	Simply write $Box(r)$ and $Box_\bfp(r)$ if
	$r_i=r$ for all $i=1\dd n$. 
	The \dt{width} of the box $B=Box_\bfp(\bfr)$ is $\bfw(B)\as 2\bfr$.
	The dimension of $B\in \intbox\RR^n$ 
	is the number $\dim(B)\in\{0\dd n\}$ of positive values in $\bfw(B)$.
	The \dt{maximum, minimum widths} of $B$ are 
	$\wmax(B)\as \max_{i=1}^n (\bfw(B))_i$,
	and similarly for $\wmin(B)$.
	\ignore{
	A set of the form
			$\cyl(h,\bfp,r)\as [0,h]\times Ball_{\bfp}(r)$
	is called a \dt{cylinder}.
	}
	The natural extension of any function $f:X\to Y$ is
	used throughout; this extension (still denoted $f$)
	is $f:2^X\to 2^Y$ where $f(X)=\set{f(x): x\in X}$
	and $2^X$ is the power set of $X$.
	If $f:\RR^n\to\RR^m$, then a \dt{box function} for
	$f$ is any function of the form
	$\intbox f:\intbox\RR^n\to\intbox \RR^m$
	that is conservative, i.e., $f(B)\ib \intbox f(B)$,
	and convergent, i.e., $\bfp=\lim_{i\to\infty} B_i$ implies
	$f(\bfp)=\lim_{i\to\infty} \intbox f(B_i)$.
	\ignore{
	Different box functions of $f$
	are distinguished by sub- or superscripts such as
	$\intbox_1 f$ or $\intbox^T f$.
	}%
	For simplicity, we
	often say ``compute $f(B)$'' but we actually compute
	$\intbox f(B)$ using interval methods.

	\dt{Normalized Taylor Coefficients}:
	Although IVP theory only requires the function
	$\bff$ in \refeq{bfx'}
	to be locally Lipschitz, practical algorithms
	need $\bff$ to be $C^k$ for moderately large $k$.
	Following
	\cite{capd:rigorousDynSys:21}, we assume $k=20$
	(a global constant) in our implementations.
	For $i=0\dd k$, the $i$th \dt{normalized Taylor coefficient}
	of $\bff$ is recursively defined by
			$\bff\supn[0](\bfx) =\bfx$ and for $i\ge 1$,
	$\bff\supn[i+1](\bfx) =
					\efrac{i} \Big( J_{\bff\supn[i-1]} \Bigcdot
						\bff\Big)(\bfx)$
	where $J_\bfg$ is the Jacobian of any 
	$\bfg=\bfg(\bfx)\in C^1(\RR^n\to\RR^n)$.
\issacArxiv{
	E.g., $\bff\supn[1] = \bff$ and
		$\bff\supn[2](\bfx) = \half (J_\bff\Bigcdot \bff)(\bfx).$
	The order $k$ Taylor expansion of $\bff(\bfx)$
	at the point $t=t_0$ is
		\beql{taylor}
		\mmatX[rcl]{
		\bfx(t_0+h)
			&=& \Big\{\sum_{i=0}^{k-1} h^i \bff\supn[i](\bfx(t_0)) \Big\}
				+ h^k \bff\supn[k](\bfx(\xi))\\
			&\in& \Big\{
				\sum_{i=0}^{k-1} h^i \bff\supn[i](\bfx(t_0)) \Big\}
				+ h^k \bff\supn[k](F)
			}
		\eeql
		where $0\le \xi-t_0 \le h$ and
		$F$ is any enclosure of $\set{\bfx(t): t\in[0,h]}$.

	Since \refeq{bfx'} is autonomous, we will normally
	assume that the initial time is $t=0$.
	Up to time scaling, we can even make $h=1$. So we can let
	$h$ be an optional parameter with default value 
	$h=1$; this is mainly to
	simplify notations but our results hold for any $h>0$. 
	}
\issacArxiv{
	Recall the set $\ivp_\bff(B_0;h)$ defined in the introduction.
	If $B_0=\set{\bfp_0}$ is a singleton,
	we simple write $\ivp_\bff(\bfp_0;h)$.
	}%
	We say $\ivp_\bff(B_0;h)$ is \dt{valid}
	if for each $\bfp_0\in B_0$,
	$\ivp_\bff(\bfp_0;h)$ is a singleton (comprising the
	unique and bounded solution $\bfx$ with $\bfx(0)=\bfp_0$).
    Since $\bff$ is usually fixed, they appear
	in subscripts and is often omitted, as in $\ivp(B_0;h)$.

	\dt{Admissibility}:
	If $F$ and $E$ are (respectively) full enclosure
	and end enclosure of $\ivp_\bff(B_0,h)$, then we call 
		$(B_0,h,F)$ an \dt{admissible  triple} and
		$(B_0,h,F,E)$ an \dt{admissible  quad}.
	We simply say ``admissible'' when it is clearly
	triple or quad.

	For closed convex sets $E, F\ib \RR^n$ and $h>0$,
	\cite[Lemma 1]{zhang-yap:ivp:25arxiv} says that
	$(E,h,F)$ is admissible provided and $E\subseteq interior(F)$
		\beql{tay}
			\sum_{i=0}^{k-1}
		 	[0,h]^i \bff^{[i]}(E)+[0,h]^k \bff^{[k]}(F)
				\subseteq F.
		 \eeql
	If, in addition, we have 
		$[0,h]^k \bff^{[k]}(F)\ib [\pm\veps]^n$,
	then we say $(E,h,F)$ is \dt{$\veps$-admissible}

\issacArxiv{
	\dt{Logarithmic norms.}  
	Let $\mu_p(A)$ denote the \dt{logarithmic norm}
	(\lognorm) of a square matrix $A\in\CC^{n\times n}$
	based on the induced matrix $p$-norm.  For specificity,
	assume $p=2$.  For
		$\bff\in C^1(\RR^n\to\RR^n)$, the
	\dt{logNorm bound} of $B\ib\RR^n$ is defined
	as $\mu_2(J_\bff(B))\as$ $
	\sup\set{\mu_2(J_\bff(\bfp): \bfp\in B}$
	where $J_\bff$ is the Jacobian matrix of $\bff$.
	}
	
	\issacArxiv{
	In this paper, in addition to the 
	convention of specifying problems as in the
	footnote for \refeq{endEnc}, we have
	a corresponding header for algorithms that solve
	the problem:
	  \beql{ttALGORITHM}
	  \mbox{\tt ALGORITHM}(\bfa;\coblue{\bfb};\cored{\bfd})\to \bfc.
	  \eeql
	This algorithm header has the problem arguments $\bfa;\coblue{\bfb}$,
	but might further have optional
	``hyperparameters'' $\cored{\bfd}$ which have
	no effect on the correctness of the
	algorithm, but users can use them to provide hints 
	that may speed up the algorithm.
}


%% file: inc/2-algoReview.tex
\ssect{Review of \endEncIVP\ Algorithm}
	Most validated IVP algorithms are based
	on two basic subroutines which we call
	\dt{Step A} and \dt{Step B}, with these headers:
	\beql{stepAB}
		\Indent[10]\mycolorbox{
			$\mmatX{\stepA_\bff(E;\cored{H=1}) &\to & (h,F)\\
				\stepB_\bff(E,h,F;\cored{\veps_0=\wmin(E)})
						&\to & E_1}$
			}
		\eeql
	where $(E,h,F)$ and $(E,h,F,E_1)$) are admissible.
	Note that $\cored{H},\cored{\veps_0}$ are optional, with
	the indicated default values.
	Starting with $E_0$, we call Step A then Step B to
	produce an admissible quad:
		\beql{stepAstepB}
			E_0 \overset{\stepA}{\looongrightarrow}
			(E_0,h_0,F_1) \overset{\stepB}{\looongrightarrow}
			(E_0,h_0,F_1,E_1).
		\eeql
	In general, by calling Step A then Step B on 
	$E_{i-1}$ ($i=0,1,\ldots$), we produce the
	the $i$th \dt{stage} represented by the
	admissible quad $(E_{i-1}, h_i, F_i,E_i)$.
	Iterating $m$ times, we organize these 
	quads into an \dt{$m$-stage scaffold} data structure
	$\stage=(\bft,\bfE,\bfF,\bfG)$ where
		\beql{EtF}
		\mmatX[ll]{
			\bft=(0=t_0<t_1<\cdots<t_m),\quad & \bfE=(E_0,E_1\dd E_m)\\
			\bfF=(F_1\dd F_m),\quad & \bfG=(G_1\dd G_m)}
		\eeql
	where $t_i=t_{i-1}+h_i$ and the $G_i$'s is 
	the $i$th \dt{refinement substructure} used
	for refining the $i$th stage.  This scaffold
	allows us to achieve much stronger objectives than are 
	possible in previous IVP algorithms.

	Let $\stage$ be an $m$-stage scaffold $\stage$.
	It is self-modified by 2 subroutines: the first is
	$\stage.\Extend$, which turns $\stage$
	into a $m+1$-stage scaffold.  It is straightforward
	(basically calling Steps A and B), so we focus on
	the second subroutine $\stage.\Refine$ with this header:
	\renewcommand{\alt}[2]{#2} 
	\renewcommand{\alt}[2]{#1} 
	\alt{
		\beql{refine}
		\hspace*{10mm}
			\includegraphics[width=0.75\columnwidth]{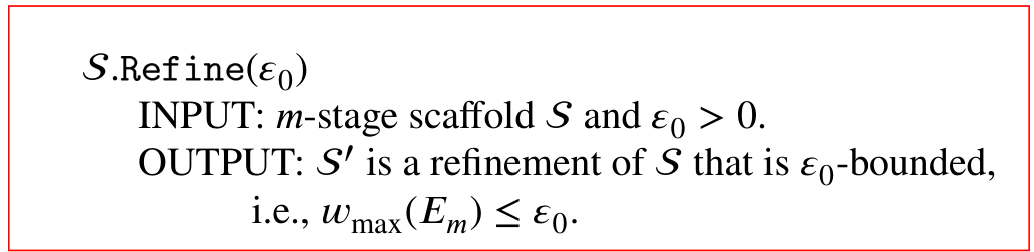}
		\eeql
	}{ 
		\Ldent\progb{
		\lline[0] $\stage.\Refine( \veps_0) $
		\lline[5] INPUT:  $m$-stage scaffold $\stage$ and
				$\veps_0>0$.
		\lline[5] OUTPUT: $\stage'$ is a refinement of
				 $\stage$ that is $\veps_0$-bounded,
		\lline[15]
				 i.e., $\wmax(E_{m}(\stage'))\le \veps_0.$}
	}
	Each iteration of the while-loop in  $\stage.\Refine$ is called a
	\dt{phase}. 
	Each phase will refine the $i$th stage of $\stage$
	(from $i=1$ to $i=m$). 
	We can visualize each phase as a row
	of the \dt{phase-stage} diagram in \refeq{phase-stage}:
	\renewcommand{\alt}[2]{#2} 
	\renewcommand{\alt}[2]{#1} 
	\alt{
		\beql{phase-stage}
		\includegraphics[width=0.7\columnwidth]{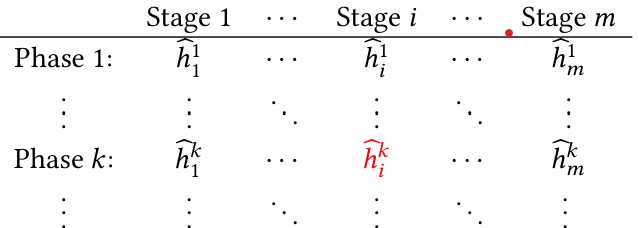}
		\eeql
	}{
			\btable{
				& Stage $1$ & $\cdots$ & Stage $i$
								& $\cdots$ & Stage $m$\\\hline
			Phase $1$: & $\whh^1_1$ & $\cdots$ &	$\whh^1_i$
								& $\cdots$ & $\whh^1_m$	\\
				$\vdots$ & $\vdots$	& $\ddots$ &$\vdots$
								& $\ddots$&	$\vdots$ \\
			Phase $k$: & $\whh^k_1$	& $\cdots$ & \cored{$\whh^k_i$}
								& $\cdots$ & $\whh^k_m$\\
				$\vdots$ & 	$\vdots$	& $\ddots$ & $\vdots$
								& $\ddots$ & $\vdots$	
			}
	}%
	Each entry of the diagram is indexed by a pair
	$(k,i)$ representing the $k$th phase at the $i$th stage.
	The operations at $(k,i)$ are controlled by the $i$th
	refinement substructure 
		\beql{Gi} G_i = G_i(\stage) =((\pi_i,\bfg_i),
				\cored{ (\ell_i, \ol\bfE_i, \ol\bfF_i)},
				(\ol\bfmu^1_i,\ol\bfmu^2_i),
				\cored{(\delta_i, h\euler_i)}).\eeql 
	Here we focus on the components in red.  We call
	$\cored{(\ell_i,\olE_i,\olF_i)}$ the
		$i$th \dt{mini-scaffold} and
	$\ell_i\ge 0$ is the \dt{level}.  It represents
	a subdivision of $[t_{i}-t_{i-1}]$
	into $2^{\ell_i}$ mini-steps of size
	$\whh_i\as (t_i-t_{i-1})2^{-\ell_i}$, and
	$\olE_i, \olF_i$ are $2^{\ell_i}$-vectors representing admissible
	quads $(\olE_i[j-1]), \whh_i, \olF_i[j],\olE_i[j])$
	(for $j=1\dd 2^{\ell_i}$.
	In \refeq{phase-stage}, we write 
	$\cored{\whh_i^k}$ for the value of $\whh_i$ in phase $k$.
	At each $(k,i)$, we can refine the $2^{\ell_i}$
	admissible quads by either calling $\stage.\eulertube(i)$
	or $\stage.\Bisect(i)$.  Ideally, we want to call
	$\stage.\eulertube(i)$ because it is very fast.
	But it has a precondition, viz., {\em $\whh_i$
	is less than $h\euler_i$ in $G_i$} (see \refeq{Gi}).
	If the precondition fails, we call $\stage.\Bisect(i)$
	and also increment the level $\ell_i$.
	This $\Bisect(i)/\eulertube(i)$ design
	is critical for the overall performance
	of our algorithm (see \cite[Table B]{zhang-yap:ivp:25arxiv}).
\issacArxiv{
	The precondition $\whh_i<h\euler_i$ guarantees that
	the polygonal Euler path produced by Euler Tube method
	is guaranteed to lie in the $\delta_i$-tube of
	each solution $\bfx\in\ivp(E_i,t_i-t_{i-1})$
	(\cite[Lemma 6, Sect.2.5]{zhang-yap:ivp:25arxiv}).
	The value $h\euler_i$ is computed by the function
		$h\euler(H,\olM,\mu,\delta)$
	in \cite[Eq.~(12)]{zhang-yap:ivp:25arxiv}.
	}%
	\ignore{
		For $k=1$, $\delta^1_i$ is $\veps_0$ if $i=m$
		and otherwise inherited from $(m-1)$-stage substructure
		before $\Extend(\veps_0,H)$.
		For $k>1$, $\delta^k_i$ halved after a call to
		\eulertube\ subroutine.
	
	If the step size less than
		\beql{olh}
				h^{\mathrm{euler}}(H,\olM,\mu,\delta)
		\eeql
	(see \cite[Eq.~(12)]{zhang-yap:ivp:25arxiv}),

	Additionally, a radical
	transform technique is applied, with the transform fixed
	per stage due to its computational cost (involving
	symbolic components).
		Empirical evidence in
	\cite{zhang-yap:ivp:25arxiv} demonstrates the
	effectiveness of these techniques.
	}%

%
%
%
%
%
%

%% file: inc/2-enhancedEE.tex
\ssect{The Enhanced \endEncIVP\ Algorithm}
	The \endEncIVP\ algorithm itself is unchanged from
	\cite{zhang-yap:ivp:25arxiv},
	except that it calls the following
	\dt{enhanced} $\Refine$ subroutine:
	
	{\scriptsize
	\Ldent\progb{
		\lline[0]  $\stage.\Refine(\veps_0,\bfp_0)$
						\dt{(enhanced version)}
		\lline[5]  INPUT:  $m$-stage scaffold $\stage$, 
				$\veps_0>0$ and point $\bfp_0\in E_0(\stage)$.
		\lline[5]  OUTPUT: $\stage$ remains an $m$-stage scaffold
		\lline[25]           that satisfies $\wmax(E_{m})\le \veps_0$.
		\lline[5]  \myhlineX[2.5]{0.5}
		\lline[5]  $r_0\ass \wmax(E_m(\stage))$.
		\lline[5]  While ($r_0> \veps_0$)
		\lline[10]    \hspace*{-1.0mm}\commenT{The original code,
		``$E_0(\stage)\ass \half E_0(\stage)$,'' is replaced by the
		next lines:}
		\lline[10]    \cored{$\displaystyle \olmu \ass
		\max\big\{\ol\bfmu^1(G_i(\stage))\;:\; i=1\dd m\big\}$.}
		\lline[10] \cored{$\Delta\ass
		\max_{i=1}^m(\delta(G_i(\stage))) $ }
		\lline[10] \cored{If ($e^{ \olmu}\,\Delta\, m < \veps_0/8$
		and  $\;e^{\olmu
		(t_m(\stage)-t_0(\stage))}\,\wmax(E_0(\stage))< \veps_0/2$)} 
		\lline[15] \cored{For ($i=1\dd m$)}
				\Commentx{CONDITION 3 \refeq{con-3} holds}
		\lline[20] \cored{$G_i(\stage).\delta\ass
		2G_i(\stage).\delta$}
		\lline[20] \cored{$\stage.\eulertube(i)$.} \Commentx{Create a
		chain}
		\lline[15] \cored{Return $\stage$ } \Commentx{Stop the
		algorithm}
		\lline[10]    $r_0\ass\wmax(E_m(\stage))$
		\lline[10]    For ($i=1\dd m$) \Commentx{Begin new phase}
		\lline[15] \cored{If (	$e^{ \olmu}\,\delta(G_i(\stage))\, m
		< \veps_0/8$)}  \Commentx{CONDITION 2 \refeq{con-2} holds}
		\lline[20] \cored{Continue} \Commentx{Skip this stage}
		\lline[15]      Let $\ell \ass G_i(\stage).\ell$ and $\Delta
		t_i \ass \bft(\stage).t_i - \bft(\stage).t_{i-1}$.
		\lline[15]      $H\ass (\Delta t_i)2^{-\ell}$.
		\lline[15]      If ($H>\wh{h}$)
		\lline[20]        $\stage.\Bisect(i)$
		\lline[15]      Else
		\lline[20]        $\stage.\SubrSeven(i)$
		\lline[20]   \issacArxiv[%
		 \ldots \Commentx{See \Refine\ in \cite{zhang-yap:ivp:25arxiv}}
		 ]{     Let $G_i(\stage)$ be
		\cored{$(\pi,\bfg,\ol\bfmu^1,\ol\bfmu^2,\delta, \wh{h},
			(\ell,\bfE, \bfF))$} after the Euler Tube.
		\lline[20]        \commenT{We must update $\delta$ and
			$\wh{h}$ for the next Euler Tube event.}
		\lline[20]        $G_i(\stage).\delta\ass \delta/2$
		\Commentx{(Updating the $\delta$ target is easy)}
		\lline[20]        \commenT{Updating $\wh{h}$ to reflect the
		 $\delta$ requires six preparatory steps: (1)--(6)}
		\lline[20]        (1)\quad $E_i(\stage)\ass \bfE[2^{\ell}]$
		\Commentx{End-enclosure for stage}
		\lline[20]        (2)\quad $F_i(\stage)\ass
		\bigcup_{j=1}^{2^\ell} \bfF[j]$
		\Commentx{Full-enclosure for stage}
		\lline[20]        (3)\quad $\mu^2_2\ass
		\max\{\ol\bfmu^2[j]\;:\; j=1\dd 2^\ell\}$
		\lline[20]        (4)\quad $\delta'_1\ass
		\TransformBound(\delta,\pi, F_i(\stage))$
		\lline[20]        (5)\quad $\ol{M}\ass
		\big\|\bfg^{[2]}(\pi(F_i(\stage)))\big\|_2$
		\lline[20]        (6)\quad $htmp \ass
		\min\big\{\Delta t_i,\; h_{\mathrm{euler}}
		(\wh{h},\ol{M},\mu^2_2,\delta'_1)\big\}$
		\lline[20]        $G_i(\stage).\wh{h}\ass htmp$
		\Commentx{Finally, update $\wh{h}$}
	}
		\lline[10]    \hspace*{-3.0mm}\Commentx{End of For-Loop}
		\lline[10]    \hspace*{-1.0mm}\commenT{The original code,
			``$E_0(\stage)\ass \half E_0(\stage)$,''
			is replaced by the next lines:}
		\lline[10]    \cogreen{If
			$e^{\olmu (t_m(\stage)-t_0(\stage))}
			\wmax(E_0(\stage))>\veps_0/2$}
					\Commentx{CONDITION 1 \refeq{con-1} holds}
		\lline[15]      \cogreen{$\displaystyle E_0(\stage)\ass
			\bfp_0+\tfrac12\bigl(E_0(\stage)-\bfp_0\bigr)$}
		\lline[10]    $r_0\ass\wmax(E_m(\stage))$
		\lline[5]     \hspace*{-3.0mm}\Commentx{End of While-Loop}
		\lline[5]     Return $\stage$
	}}
	
	The optional argument $\bfp_0$ in \refeq{endEnc}
	is passed to the enhance \Refine.
	In the \Refine\ of \cite{zhang-yap:ivp:25arxiv}, \(E_0(\stage)\) is
	updated by the simple line
	\beql{E0old}
	E_0(\stage)\ass \tfrac12 E_0(\stage)
	\eeql
	to halve \(E_0(\stage)\) at the end of the while-loop. We now shrink
	\(E_0(\stage)\)
	conditionally, as indicated by the line in green. Moreover, we
	shrink
	\(E_0(\stage)\) toward the point \(\bfp_0\); the shrinkage step
	\refeq{E0old} is now
	\beql{shrink}
	E_0(\stage)\ass \bfp_0 + \tfrac12\bigl(E_0-\bfp_0\bigr).
	\eeql
	
	Note that ``\(E_0-\bfp_0\)'' shifts \(E_0\) so that \(\bfp_0\)
	is at origin;
	after halving \(\tfrac12(E_0-\bfp_0)\), we translate the origin of
	\(\tfrac12(E_0-\bfp_0)\) back to \(\bfp_0\).

	The enhancement amounts to trying to avoid
	unnecessary shrinkage of the initial enclosure $E_0$.
	This is estimated as follows:
	recall that a \dt{chain} is a sequence
		$
		C=(1\le k(1)\le k(2)\le \cdots \le k(m)),
		$
	such that \SubrSeven\ is invoked
	at phase-stage $(k(i),i)$ for each $i$.
	It is shown in \cite[Appendix A, (H5)]{zhang-yap:ivp:25arxiv}
	that
		\beql{chain}
		r_m(C) \;\le\; e^{ \olmu}\bigl(r_0(C)+\Delta(C)\,m\bigr)
		\;=\;
		\underbrace{e^{ \olmu}r_0(C)}_{\text{$r_0$-part}}
		\;+\;
		\underbrace{e^{ \olmu}\Delta(C)\,m}_{\Delta\text{-part}},
		\eeql
	where $r_m(C)$ and $r_0(C)$ are one half of the maximal widths of the
	last and first
	end enclosures at the corresponding phases, respectively;
	$\olmu$ is the maximal logarithmic norm bound from the first phase;
	and $\Delta(C)=\max_{i=1}^m \delta(G_i(\stage))$ is evaluated at the
	initial phase of $C$.
	For a given chain $C$, if 
	\(2r_m(C)<\veps_0\) then the \Refine\ procedure 
	halts.  
	Hence, by \refeq{chain}, \Refine\ will terminate as soon as both the
	\(r_0\)-part and the \(\Delta\)-part are bounded by \(\veps_0/4\);
	i.e., when
		\beql{cond12}
		e^{ \olmu}r_0(C)<\frac{\veps_0}{4}
		\qquad\text{and}\qquad
		e^{ \olmu}\Delta(C)\,m<\frac{\veps_0}{4}.
		\eeql
	To implement this logic in the Enhanced \Refine,
	we specify three conditions:
	
	{\footnotesize
	\begin{align}
		\text{(CONDITION 1)} &&
			\Big(\half \wmax(E_0(\stage))e^{ \olmu}  < \veps_0/4\Big)
							\label{eq:con-1} \\
		\text{(CONDITION 2)} &&
			\Big(e^{ \olmu}\delta(G_i(\stage)) m < \veps_0/8 \Big)
							\label{eq:con-2} \\
		\text{(CONDITION 3)} &&
			\Big(e^{ \olmu}\,\Delta\, m < \veps_0/8\Big)
			\text{ and }
			\Big(\half \wmax(E_0(\stage))e^{ \olmu}  < \veps_0/4\Big)
							\label{eq:con-3} \\
			&& \text{where $\Delta\as \max_{i=1}^m \delta(G_i(\stage))$}
						\nonumber
	\end{align}
	}
	In the \Refine\ code, we test for these conditions and
	take corresponding actions:
	\bitem
		\item If CONDITION 1 holds, we do not shrink \(E_0(\stage)\).
		\item If CONDITION 2 holds, we skip skip stage $i$.
		\item If CONDITION 3 holds, for all $i=1\dd m$, we
			 double value of $\delta(G_i(\stage))$
			 and invoke \(\eulertube\), and exit \Refine.
	\eitem

	Note that \refeq{con-2} in CONDITION~2 requires $<\veps_0/8$,
	rather than $<\veps_0/4$ as in \refeq{cond12}.
	This is because in $\stage.\Refine$
	we always call $\eulertube$ first and then halve $\delta$.
	Consequently, when CONDITION~2 is detected to hold, the corresponding
	stage has just used twice the current value of $\delta$ and invoked
	$\eulertube$.
	
	\ignore{
	Since stages satisfying CONDITION~2 are skipped, and CONDITION~3
	requires that all stages satisfy CONDITION~2.  Since
		$\Delta $
	The requirement of CONDITION~3 can then be written as
		$ e^{\bar\mu}\,\Delta\, m < \veps_0/8, $
	which implies that all stages are in the state obtained after calling
	$\eulertube$ with the doubled $\delta$.
	Therefore, doubling the values $\delta\bigl(G_i(\stage)\bigr)$ and
	invoking
	$\eulertube$ for every stage produces a new chain that satisfies
	\refeq{cond12}.
	}%

%% file: inc/2-endCoverAlgo.tex
\ssectL[endCover]{The End Cover Algorithm}
	We construct 
	an $\veps$-cover of $\End_{\bff}(B_0,H)$ by invoking 
	\texttt{EndEnc}, which returns a sequence of box 
	pairs $\{(\underline{B}_i,\overline{B}_i)\}_{i=1}^N$.
  
	For the given initial box $B_0$, 
	if 
	$
	B_0 \subseteq \bigcup_{i=1}^m \ulB_i,
	$
	then the union of the $\olB_i$,namely
	$
	S \as \bigcup_{i=1}^m \olB_i,
	$
	constitutes a valid enclosure; in other words,
	\(\EndCover_\bff(B_0,\veps,H) \to S\).
	
	{\scriptsize
	\Ldent\progb{
		\lline[0]
		$\EndCover_\bff(B_0,\veps,H) \to S$
		\lline \myhlineY
		\lline[0] INPUT:
		\(B_0\ib\RR^n\), \(\veps>0\),$H>0$.
		\lline[0] OUTPUT:
		\(S\ib\RR^n\) satisfying
		\lline[10]
		$
		\End_\bff(B_0,h)
		~\ib~ S ~\ib~
		\Big(\End_\bff(B_0,h) \oplus [\pm \veps]^n\Big).
		$
		\lline
		\lline[5] $S_0\ass \{B_0\}$, $S\ass \emptyset$,
		$P\ass \emptyset$
		\lline[5] While $S_0\ne \emptyset$
		
		\lline[10]  Let $B.pop(S_0)=\prod_{i=1}^{n}[a_i,b_i]$
		\lline[10] $\bfp=mid(B)$ \Commentx{Midpoint of $B$}
		\lline[10] $(\ulB,\olB)\ass \endEnc(B,\veps; \bfp,H)$
		\lline[10] If($\ulB\ne B$)
		\lline[15] $I_i^{k_i}\as \left[ a_i + \frac{k_i}{2}(b_i - a_i), \; a_i + \frac{k_i+1}{2}(b_i - a_i) \right]$
		\lline[15] $S_0\ass S_0\cup 
		\left\{ \prod_{i=1}^n I_i^{k_i}: k_i \in \{0, 1\} \right\}$.
		\lline[10] $S\ass S\cup \olB$.
		\lline[5] Return $S$.
	}
}


%% file: inc/3.tex
\sectL[cover]{Boundary Cover Problem}
		We now consider another approach to
		the End Cover Problem, by reducing
		it to a lower dimensional problem called
		\dt{Boundary Cover Problem}.
		\issacArxiv{
			To provide the theoretical justification, we must
			introduce the concept of a ``trajectory bundle''.
		}


\issacArxiv{
\input{inc/3-bundle}
\input{inc/3-boundaryCover}
}

%% file: inc/3-bundle.tex
\ssectL[traj]{Trajectory Bundles}
	We now use a more geometric language and call
	each $\bfx\in \ivp_\bff(\bfp_0,h)$ a trajectory.
	For any point $\bfp_0$, the classic Picard-Lindel\"of Theorem
	\cite{hairer+2:ode1:bk,hubbard-west:bk1}
	guarantees unique-existence of a single trajectory
	in $\ivp_\bff(\bfp_0,h)$, for $h>0$ small enough.
	For our validated algorithms, we must generalize
	point trajectories to ``trajectory bundles''.

	A \dt{trajectory} with \dt{time-span} $[0,h]$ 
	is a $C^1$-function of the form $\bfphi: [0,h]\to\RR^n$.
	Moreover, $\bfphi$ is a \dt{\dt{$\bff$}-trajectory} if
	$\bfphi\in \ivp(B,h)$ for some $B\ib\RR^n$
	(i.e., $\bfphi'(t) = \bff(\bfphi(t))$ and $\bfphi(0)\in B$).
	\ignore{
		for some for all $\tau\in [0,h]$, we have
			\beql{bfftraj}
			\bfphi'(\tau) \as
				\frac{d\bfphi}{dt}(\tau)= \bff(\bfphi(\tau)).\eeql
		More precisely, if $\bfphi=(\phi_1\dd \phi_n)$
		where $\phi_i:[0,h]\to\RR$ 
		then $\bfphi'(\tau) = (\phi_1'(\tau)\dd\phi_n'(\tau))$
		where $\phi'_1=\frac{d\phi_i}{dt}$.
		Then \refeq{bfftraj} says that
			$\bfphi'(\tau) = \bff(\phi_1(\tau)\dd \phi_n(\tau))$.
	}%
	Call
		\beql{trajbundle}
			\Phi: [0,h]\times B\to\RR^n\eeql
	a \dt{trajectory bundle} if $\Phi(\tau,\bfp)$
	is a continuous function on $[0,h]\times B$
	such that $\Phi(0,\bfp)=\bfp$ (for all $\bfp\in B$)
	and satisfies the following 
	\dt{stratification property}:
		\beql{stratify}
			\Big(\forall \bfp\ne\bfq\in B\Big)
			\Big(\forall \tau\in[0,h]\Big)\quad\Big[
				\Phi(\tau,\bfp)\ne \Phi(\tau,\bfq)\Big].
		\eeql
	For all $\tau\in [0,h]$, $\bfp\in B$, we have the
	induced functions $\Phi|_\bfp$ and $\Phi|^\tau$ 
	\beqarryl[ll]{bfp-traj}
		\Phi|_\bfp: [0,h]\to\RR^n 
				& \text{($\bfp$-trajectory)},\\
		\Phi|^\tau: B\to\RR^n 
				& \text{($\tau$-end map)}.\\
	\eeqarryl
	where $\Phi|_\bfp(\tau)=\Phi|^\tau(\bfp)=\Phi(\tau,\bfp)$.
	Call $\Phi$ a \dt{$\bff$-bundle} on $(B,h)$ 
	If each $\Phi|_\bfp$ is a $\bff$-trajectory ($\bfp\in B$),
	then denote this bundle by $\Phi = \Phi_{\bff,B,h}$.
	\issacArxiv{
	The stratification property \refeq{stratify} says that
	the graphs of the trajectories $\Phi|_\bfp$
	and $\Phi|_\bfq$ for $p\ne q$ are pairwise disjoint.
	}
	Before stating our bundle theorem, we need a useful lemma.
	Define $\valid_\bff(H)$ to be the set of all points
	$\bfp\in\RR^n$ such that $\ivp_\bff(\bfp,H)$ is valid.

	\bleml[valid]
		The set $\valid_\bff(H)$ is an open set for any $H>0$.
	\eleml

	This lemma is trivial when
	$\valid_\bff(H)$ is empty since empty sets are open.
	Here is the bundle
	version the classic Picard-Lindel\"of Theorem.

	\bthmT[bundle]{Global Picard-Lindel\"of Bundle} \ \\
		If $\bff:\RR^n\to\RR^n$ is locally Lipschitz, and
		$B\ib \valid_\bff(H)$, then
		the $\bff$-bundle $\Phi_\bff(B,H)$ exists.
	\ethmT

	The bundle in this theorem is ``global'' because $H>0$ is 
	specified in advance.
	In contrast, the standard Picard-Lindel\"of theorem
	only gives unique-existence of a ``local'' trajectory, i.e.,
	for sufficiently small $h>0$.

	\dt{The End Map:}
	If $\Phi$ is the trajectory bundle in \refeq{trajbundle},
	the \dt{$\tau$-slice} (for $\tau\in [0,h]$) of $\Phi$ is the set
		\beql{phitau}
			\Phi|^\tau\as \set{\Phi(\tau,\bfp): \bfp\in B}.\eeql
	E.g., the $0$-slice is just $B$.  The $h$-slice
	$\Phi|^h$ is called the \dt{end-slice}.
	\ignore{
	I.e., the graph of $\Phi$,
		$$gr(\Phi)=\set{(t,\bfp,\Phi(t,\bfp)):
					t\in[0,h], \bfp\in B}$$
	can be partitioned into the (marked) graphs
	of the individual trajectories $gr\big(\Phi|_\bfp\big)$
	for $\bfp\in B$.
	}%
	Define the \dt{end map} of $\Phi$ to be
		\beql{endmap}
			\End_\Phi: B\to\RR^n
		\eeql
	where $\End_\Phi(\bfp)= \Phi(h,\bfp)$.  
	\bthml[homeo]
		Let $\Phi:[0,h]\times B\to\RR^n$ be a trajectory
		bundle where $B\in\intbox\RR^n$.
		Then the end map $\End_\Phi$ of \refeq{endmap}
		is a homeomorphism from $B$ onto the end-slice $\Phi|^h$.
	\ethml

	Note that our theorem holds even when $\dim(B)<n$; we
	need this generality in our boundary technique below.
	The proof of this theorem depends on
	a famous result of L.E.J.~Brouwer (1912):
	\cheeX{Bw, please obtain the reference for Brouwer, to cite?}
	\bwX{Brouwer L.E.J. Beweis der Invarianz des 
		{$\displaystyle n$}-dimensionalen Gebiets,
		Mathematische Annalen 71 (1912),
		pages 305–315; see also 72 (1912), pages 55–56
		}

	\bproT[invariance]{Brouwer's Invariance of Domain}\ \\
		If $U$ is an open set in $\RR^n$ and
		$f:U\to\RR^n$ is continuous and injective,
		then $f(U)$ is also open.
		In particular, $U$ and $f(U)$ are homeomorphic.
	\eproT


	\cheeX{Bingwei, please check this
	\myHref{https://en.wikipedia.org/wiki/Invariance\_of\_domain}{wiki
		article}.  Our application needs $U\ib\RR^n$ to be
		a closed set and $\dim(U)$ may be less than $n$.
		It is possible there there are generalizations
		where $U$ is relatively open in $\RR^n$  (I thought I have
		seen this in my literature search).
	}

\bwX{ 
	We can also apply the result from
	\myHref{https://proofwiki.org/wiki/Continuous_Bijection_from_Compact_to_Hausdorff_is_Homeomorphism
	}
	{Continuous Bijection from Compact to Hausdorff is Homeomorphism}
	which states that a continuous bijection from a compact space to a
	Hausdorff space is a homeomorphism.}

	\ignore{To exploit the Invariance
	of Domain theorem, we henceforth consider
	trajectory bundles $\Phi$ with cylinder domains,
	$\Phi: \cyl(h,\bfp,\bfr)\to \RR^n$ (see \refSSec{notation}).
	}

%% file: inc/3-boundaryCover.tex
\ssectL[boundary]{Reduction of End Cover to Boundary Cover}

	\refThm{homeo} justifies the following
	\dt{Boundary Algorithm} for computing
	an $\veps$-box cover of $\End_\bff(B,h)$:
	if $B\in\intbox \RR^n$, we compute
	an $\veps$-box cover $\calC_i$ ($i=1\dd 2n$)
	for each of the $2n$ faces of $B$. Then
		$$\ol\calC \as \bigcup_{i=1}^{2n} \calC_i$$
	is an $\veps$-box cover for the boundary
	$\partial(\End_\bff(B,h))$.
	Next, we compute a \dt{filler cover} $\calC_0$ to
	cover the ``interior'' of $\bigcup \ol\calC$ so that
		$\ol\calC \cup \calC_0$
	forms an $\veps$-box cover of $\End_\bff(B,h)$.
	
	Thus, the Boundary Algorithm reduces an
	$n$-dimensional cover problem to several
	$(n-1)$-dimensional ones, modulo
	the \dt{Fill Problem} of computing $\calC_0$;
	the header for this problem is thus
		$$\filler(\ol\calC,\veps)\ssa \calC_0$$
	Unfortunately, this appears to be an ill-formed problem
	unless $\veps$ is small enough.

	To understand the difficulty, note that $\bigcup\ol\calC$
	is a connected set. The boundary of $\bigcup\ol\calC$
	is the union of finitely many components,
	$S_0\cup S_1\cup\cdots\cup S_k$ where each $S_j$
	is topologically an $(n-1)$-sphere.
	Moreover, $S_0$ is the unique exterior component such that
	the other $S_i$'s ($i=1\dd k$) lies in the interior of $S_0$.
	Also the $S_i$'s are not nested (if $S_i$ is contained
	in some $S_j$ then $j=0$).  We want the filler set
	$\calC_0$ to lie inside $S_0$, and to cover up
	{\em some} of these $S_i$'s.  But identifying the $S_i$'s to
	cover is a non-trivial problem especially in higher dimensions. 
	Let us say that $\veps$ is \dt{small enough} if $k=1$.
	If $\veps$ is small enough, then the boundary
	$\partial(\bigcup \ol\calC)$ has two components $S_0, S_1$

	with $S_1$ inside $S_0$.  The boxes of $\calC_0$
	just have to cover the interior of $S_1$ but stay within $S_0$. 
	This is a relatively simpler
	problem of computational geometry, with no need
	for any IVP solving. 
	In our examples ($n=2$), we can visually
	see that our $\veps$ is small enough.
	In our experimental section, we will
	compare this Boundary Algorithm with 
	$\EndCover_\bff(B_0,\veps)$.

%% file: inc/4.tex
\sectL[complexity]{Complexity Analysis}
	This section provides a complexity analysis of our 
	\EndCover\ algorithm in \refSSec{endCover}.
	Two preliminary remarks are necessary:
	\\ (C1)
	One might expect the complexity of \EndCover\ to be a function of
	the input instance $(E_0,\veps,H)$, and also $\bff$.
	But we do this indirectly, via two derived\footnote{
		See \cite[Lemma~7]{zhang-yap:ivp:25arxiv}.
		To be consistent with Lemma 7,
		we use the more general error bound
			$\bfveps=(\veps_1\dd \veps_n)$
		instead of $\veps>0$.
	} parameters $\olh$ and $\olB$ defined as follows:
	\beqarrayl{olB}
	\olB = \olB(E_0,H,\bfveps) &\as&
		\sum_{i=0}^{k-1} [0, H]^i \bff^{[i]}(E_0)
		+ Box(\bfveps),\\
	\olh = \olh(E_0,H,\bfveps) &\as&
		\min\set{H,\min_{i=1}^n
			\Big(\tfrac{\veps_i}{M_i}\Big)^{1/k} }, \label{eq:olh}\\
	&& \text{where} \notag\\
		M_i &\as&
		\sup_{\bfp \in \ol{B}(E_0,H,\bfveps)}
		\Big| \big(\bff^{[k]}(\bfp) \big)_i \Big|,
		 (i=1\dd n), \label{eq:M_i} 
	\eeqarrayl
	\ (C2)
	Second, our complexity counts the number of \dt{steps},
	defined to be either assignments (e.g., $x\ass F(y)$)
	or if-tests (e.g., ``If $(H>\wh{h})$'')
	in our pseudo-codes for
	\Extend, \Refine, \SubrSeven, \Bisect, etc.
	The \dt{non-steps} in pseudo-codes are the
	subroutine calls as well as while- and for-statements.
	The overall number of steps can be bounded if we
	bound the number of iterations in while- and for-loops.
	Each step takes $O(1)$ time
	for fixed $\bff,E_0,\veps,H$.  But some
	assignments appear violate this. 
	\issacArxiv{ E.g. lines (2) and (3)
	in $\Refine$ appear to take $\Omega(2^\ell)$ steps.
	But it is not hard to re-factor our code
	so that they are truly $O(1)$.}
	
	In principle, our step complexity could be translated into
	the standard bit-complexity model based on Turing machines,
	more or less routinely.

\input{inc/4-complexity}


%% file: inc/4-complexity.tex
\ssectL[refine]{Complexity of End Enclosure}
	We first analyze the complexity of the
	\endEnclAlgo, since it is called by the \endCovAlgo.
	There are 3 components:
	\begin{enumerate}
		\item The total number of calls to \Extend\ and \Refine,
		i.e., the number of stages generated by the
		\endEnc\ algorithm;
		\item The complexity of a single call to \Extend;
		\item The complexity of a single call to \Refine.
	\end{enumerate}
	
\sssect{Bound on the number of stages}
	We first bound the 
	number of stages produced by the \texttt{EndEnc} 
	algorithm:
	\bleml[Numberofstages]
	\ \\The number of stages in
		\endEnc$(E_0,\bfveps; \bfp,H)$ is at most
		$1+\floor{H/\olh}$ where
		$
		\olh=\olh(\olE,H,\bfveps)$, see \refeq{olh}
		and 
		$
		\olE =\Image_\bff(E_0,H)+Box(\bfveps).
		$
	\eleml
\sssect{Complexity of \Extend}
	The subroutine $\stage.\Extend$ appends a new
	stage to scaffold $\stage$. 
	This is accomplished by one call each of \stepA\ and \stepB,
	together with some auxiliary operations, in order to construct
	admissible triples and validated enclosures.
	All these are loop-free operations except for \stepA. 
	\issacArxiv{
	E.g., \stepB\ basically performs Taylor
	expansions of order $k$ for an $n$-dimensional system,
	and counts as $O(1)$ steps.}

	\blemT[complexityExtend]{Extend}
	Given a scaffold \(\stage\) with \(m\) stages and parameters
	\(\bfveps>0\) and \(H>0\), the subroutine \stepA\ executed within
	\(\stage.\Extend(\bfveps,H)\) performs at most
	$$
	\left\lceil \log_2\!\left(\frac{H}{\olh}\right) \right\rceil
	$$
	iterations, where
	\(\olh=\olh(E_m(\stage),H,\bfveps)\) is defined in \refeq{olh}
	\elemT

\sssect{Complexity of \Refine}
	The complexity of \Refine\ is $O(1+N_0)$ where $N_0$
	is the maximum of $N_1$ and $N_2$ where
	$N_1$ is the number of phases needed
	to satisfy condition~\refeq{con-1},
	and
	$N_2$ is the number of phases needed to satisfy
	condition~\refeq{con-2} for every stage.
	After $N_0$ phases, we need one more phase in which
	every stage will call \eulertube\ to form a chain.
	By~\refeq{chain}, \Refine\ terminates.
	The ``$1+$'' refers to this last phase.
	
	\blemT[complexityRefine]{\Refine}
	Let $\stage$ be an $m$-stage scaffold, and $\veps_0>0$.  
	For each stage $i$ write
	$\Delta t_i = t_{i+1}(\stage)-t_i(\stage)$ and
	\[
	G_i(\stage) = ((\pi_i, \bfg_i),
				(\ell_i, \bfE_i, \bfF_i),
				(\ol\bfmu^1_i, \ol\bfmu^2_i), 
				(\delta_i, \wh{h}_i)).
	\]
	The number of phases in $\stage.\Refine(\veps_0)$ is at most
	$1+N_0$ where
		\bitem
		\item $N_0 \as \max\set{N_1,N_2}$,
		\item $N_1 \as \Bigl\lceil \log_2
			\frac{2 \, \wmax(E_0(\stage)) \, e^{\olmu}}{\veps_0}
					\Bigr\rceil$,
		\item $N_2 \as \#\text{EulerTube} + \#\text{Bisection}$
		\item $\#\text{EulerTube} \as \Bigl\lceil \log_2
			\frac{8 m \, \Delta_{\max}
			\, e^{\olmu}}{\veps_0} \Bigr\rceil$
		\item $\#\text{Bisection} \as 
				\Biggl\lceil 
					\displaystyle{\max_{i=1,\dots,m, p\in I_i}} \left(
						\log_2 \Biggl(
				\frac{\Delta t_i}{\, h^{\euler} \Bigl( \Delta t_i, M_i, 
					p, 
					\delta_i'
					\Bigr) }
				\Biggr)\right) \Biggr\rceil$\\
		\item Moreover,
			$I_i= \big[
				\displaystyle\min_{\bfq\in \pi(F_i(\stage))}
					(\mu_2(J_{\bfg_i}(\bfq))),
				\displaystyle\max_{i=1\dd m}(max \ol\bfmu_i^2)\big]$,
		\item 	$ \olmu \as \displaystyle \max_{i=1\dd m}(max \ol\bfmu_i^1)$, 
				$\Delta_{\max}  \as  \displaystyle\max_{i=1,\dots,m} \delta_i$,
		\item 	$M_i \as \|\bfg_i^{[2]}(\pi(F_i(\stage)))\|_2$,

		\item $\delta_i'\as \TransformBound
			\left(\min\!  \left( \delta_i,
				\frac{\veps_0}{16 m e^{\olmu}}\right),
				\pi,F_i(\stage)\right).$
		\eitem
	\elemT
	
\sssect{Complexity of \endEnc}
	By \refLem{Numberofstages},
	\endEnc\ will generate $\le 1+\lfloor H/\olh\rfloor$ stages.
	Within each stage, the \Extend\ and \Refine\ 
	subroutines are called once, and their complexity 
	are bounded as in
	\refLem{complexityExtend} and \refLem{complexityRefine}.
	It follows that the step complexity of \endEnc\ is 
		\beql{complexityEndInc}
			O\Big((H/\olh)\cdot
				\big((\log_2(H/\olh)
				+ N_0\big)\Big)
		\eeql
	
\ssect{Complexity of \EndCover}
	We return from the analysis of \endEnc\ to our
	\EndCover\ algorithm. This higher-level procedure 
	orchestrates multiple calls to \endEnc, and 
	understanding its complexity requires integrating the 
	insights from the subroutine analyses.
	
	To complete the picture, we now bound the number of 
	times \EndCover\ invokes \endEnc. The following 
	theorem addresses this:
	
	\bthml[complexityEndCover]
	Let $\olB = \image(\ivp(B_0,H)) + Box(\veps)$ 
	and
	$\olmu = \mu_2(J_{\bff}(\olB))$.
	Then $\EndCover(B_0,H,\veps)$ will call \endEnc\ at most
	\issacArxiv[$
	\Theta\!\left(
	\left(\frac{2\wmax(B_0)e^{\olmu}}{\veps}\right)^n
	\right)
	$]{	$$
		\Theta\!\left(
		\left(\frac{2\wmax(B_0)e^{\olmu}}{\veps}\right)^n
		\right)
		$$}
	times.
	\ethml

%% file: inc/5.tex
\sect{Experiments}
	This section presents empirical evaluations to 
	demonstrate the practical efficacy and advantages of 
	our proposed algorithms.  Our experiments are based on the
	ODE models given in \refTab{problems}.
	Our open source C++ programs (include an executable),
	data, Makefile are available in our project
	home \cite{core:home-ivp}.
	All timings are taken on a laptop with a 
	13th Gen Intel Core i7-13700HX 2.10 GHz processor and 16.0 GB RAM.	

	\dt{LIMITATION:} Our current implementation
	are based machine-precision arithmetics.  
	As explained in \refSSec{literature}, in principle, we
	could achieve rigorous arbitrary precision implementation.
	This is future plan.

\newcommand*{\myalign}[2]{\multicolumn{1}{#1}{#2}}
		\begin{table*}[] \centering
		{\tiny
			\btable[l| l | l | l | l | l ]{
				Eg* & \myalign{c|}{\dt{Name}}
				& \myalign{c|}{$\bff(\bfx)$}
				& \myalign{c|}{\dt{Parameters}}
				& \myalign{c|}{\dt{Box} $B_0$}
				& \myalign{c}{\dt{Reference}}
				\\[1mm] \hline \hline
				Eg1 & Lotka–Volterra Predator-Prey
				& $(\phantom{-}ax(1-y), -by(1-x))$
				&  $(a,b)=(2,1)$
				& $Box_{(1,3)}( 0.1)$
				& \cite{moore:diffEqn:09},
				\cite[p.13]{bunger:taylorODE:20}
				\\[2mm] \hline 
				Eg2 & Van der Pol 
				& $(y, -c(1-x^2)y -x)$
				& $c=1$
				& $Box_{(-3,3)}( 0.1)$
				& \cite[p.2]{bunger:taylorODE:20}
				\\[1mm] \hline
				Eg3 & Asymptote
				& $(x^2, -y^2 + 	7x)$
				& N/A
				& $ Box_{(-1.5,8.5)}(0.01 )$
				& N/A
				\\[1mm] \hline
				Eg4 & Quadratic
				& $(y,x^2)$
				& N/A
				& $ Box_{(1,-1)}(0.05)$
				& \cite[p.11]{bunger:taylorODE:20}
				\\[1mm] \hline
				Eg5 & FitzHugh--Nagumo
				& $(x - \frac{x^3}{3} - y + I,
				\varepsilon (x + a - b y))$
				& $(a,b,\varepsilon, I)=(0.7, 0.8, 0.08, 0.5)$
				& $Box_{(1,0)}(0.1)$
				& \cite{sherwood:fitzhugh-naguma:13}
				\\[2mm] \hline
				Eg6 & Reduced Robertson (2D)
				& $(-k_1\,x + k_2\,y(1-x-y),
				\phantom{-}k_1\,x - k_2\,y(1-x-y) - k_3\, y^2)$
				& $(k_1,k_2,k_3)=(0.04,10^4,3\times 10^7)
				$
				& $Box_{(1,0)}(10^{-6})$
				& \cite{hairer:ode2:bk}
				\\[2mm] \hline
				Eg7 & Lorenz
				& $(\sigma(y-x), x(\rho-z)-y, xy-\beta z)
				$
				&
				$(\sigma,\rho,\beta)=(10,28,8/3)$
				& $ Box_{(15,15,36)}(0.001)$
				& \cite[p.11]{bunger:taylorODE:20}
				\\[1mm] \hline
				Eg8 & R\"ossler
				& $(-y-z,x+ay,b+z(x-c))$
				&
				$(a,b,c)=(0.2,0.2,5.7)$
				& $ Box_{(1,2,3)}(0.1)$
				& \cite{rossler:chaos:76}
				\\[1mm] \hline
				
		} }
		\caption{List of IVP Problems}
		\label{tab:problems}
	\end{table*}

\input{inc/5-experiments}


%% file: inc/5-experiments.tex
\ssect{Comparison with other validated Solvers}
\issacArxiv{
	\begin{table*}
		\centering
		{\small
			\begin{tabular}{c|c|c|c|c|c}
				\hline
				\textbf{Case} & \textbf{Method} & $T$
				& $B_1 = (\mathrm{mid}) \pm (\mathrm{rad})$
				& $\frac{\wmax(others)}{\wmax(ours)}$
				& \textbf{Time (s)} \\
				\hline
				\multirow{15}{*}{Eg1} & Ours & \multirow{5}{*}{2.00}
				& $(0.08,0.57)\pm(0.02,0.02)$ & 1 & $0.02|0.05$ \\
				& \simpleIVPdirect\ & & $(0.08,0.58)\pm(0.16,0.13)$
				& 6.66 & 0.52 \\
				& VNODE-LP & & $(0.08,0.58)\pm(0.04,0.05)$ & 2.26 & 0.17
				\\
				& CAPD & & $(0.08,0.58)\pm(0.05,0.05)$ & 2.12 & 0.01 \\
				\cline{2-6}
				& Ours & \multirow{5}{*}{4.00} &
				$(1.46,0.19)\pm(0.40,0.05)$
				& 1 & $0.13|0.08$ \\
				& \simpleIVPdirect\ & & \cored{Timeout} & N/A &
				\cored{Timeout} \\
				& VNODE-LP & & $(1.43,0.34)\pm(3.23,0.91)$ & 8.07 & 0.39
				\\
				& CAPD & & $(1.45,0.19)\pm(3.38,0.88)$ & 8.35 & 0.02 \\
				\cline{2-6}
				& Ours & \multirow{5}{*}{5.50} &
				$(1.14,2.95)\pm(0.67,0.49)$ & 1 & $3.42|1.21$ \\
				& \simpleIVPdirect\ & & \cored{Timeout} & N/A &
				\cored{Timeout} \\
				& VNODE-LP & & \cored{Invalid} &
						N/A &
				\cored{Invalid} \\
				& CAPD & & \cored{No Output} & N/A & \cored{No Output} \\
				\hline
				\multirow{10}{*}{Eg2} & Ours & \multirow{5}{*}{1.00} &
				$(-2.13,0.57)\pm(0.25,0.21)$ & 1 & $0.02|0.04$ \\
				& \simpleIVPdirect\ & & $(-2.13,0.57)\pm(0.26,0.23)$ &
				1.03 & 0.51 \\
				& VNODE-LP & & $(-2.15,0.58)\pm(0.90,3.49)$ & 13.95 &
				0.17 \\
				& CAPD & & $(-2.13,0.56)\pm(0.29,0.28)$ & 1.15 & 0.02 \\
				\cline{2-6}
				& Ours & \multirow{5}{*}{2.00} &
				$(-1.41,0.96)\pm(0.38,0.35)$ & 1 & $0.77|0.14 $\\
				& \simpleIVPdirect\ & & \cored{Timeout} & N/A &
				\cored{Timeout} \\
				& VNODE-LP & &
				\cored{Invalid}&
				N/A & \cored{Invalid} \\
				& CAPD & & \cored{No Output} & N/A & \cored{No Output} \\
				\hline
				\multirow{5}{*}{Eg3} & Ours & \multirow{5}{*}{1.00} &
				$(-0.60,-6.69)\pm(0.00,0.18)$ & 1 & $0.02|0.07$ \\
				& \simpleIVPdirect\ & & $(-0.60,-6.69)\pm(0.01,0.19)$ &
				1.01 & 4.11 \\
				& VNODE-LP & & $(-0.60,-6.69)\pm(0.00,0.19)$ & 1.05 &
				0.20 \\
				& CAPD & & $(-0.60,-6.69)\pm(0.01,0.19)$ & 1.01 & 0.02 \\
				\hline
				\multirow{10}{*}{Eg4} & Ours & \multirow{5}{*}{1.00} &
				$(0.28,-0.58)\pm(0.15,0.16)$ & 1 & $0.01|0.02$ \\
				& \simpleIVPdirect\ & & $(0.28,-0.59)\pm(0.16,0.17)$ &
				1.06 & 0.01 \\
				& VNODE-LP & & $(0.28,-0.59)\pm(0.16,0.18)$ & 1.10 & 0.03
				\\
				& CAPD & & $(0.28,-0.59)\pm(0.16,0.17)$ & 1.06 & 0.01 \\
				\cline{2-6}
				& Ours & \multirow{5}{*}{4.00} &
				$(-0.65,0.25)\pm(0.24,0.57)$ & 1 & $0.12|0.06$ \\
				& \simpleIVPdirect\ & & $(-0.68,0.34)\pm(1.01,2.12)$ &
				3.71 & 0.01 \\
				& VNODE-LP & & $(0.59,2.53)\pm(22.57,88.77)$ & 155.74 &
				0.09 \\
				& CAPD & & $(-0.68,0.34)\pm(7.51,25.53)$ & 44.79 & 0.01 \\
				\hline
				\multirow{10}{*}{Eg5} & Ours & \multirow{5}{*}{1.00} &
				$(1.76,0.17)\pm(0.11,0.10)$ & 1 & $0.03|0.07$ \\
				& \simpleIVPdirect\ & & $(1.76,0.17)\pm(0.17,0.10)$ &
				1.55 & 0.01 \\
				& VNODE-LP & & $(1.76,0.17)\pm(0.10,0.10)$ &
							\coblue{0.95} & 0.09
				\\
				& CAPD & & $(1.76,0.17)\pm(0.18,0.10)$ & 1.66 & 0.02 \\
				\cline{2-6}
				& Ours & \multirow{5}{*}{4.00} &
				$(1.68,0.68)\pm(0.08,0.10)$ & 1 & $1.22|0.13$ \\
				& \simpleIVPdirect\ & & \cored{Timeout} & N/A &
				\cored{Timeout} \\
				& VNODE-LP & & $(1.68,0.68)\pm(0.07,0.11)$ & 1.13 & 0.41 \\
				& CAPD & & \cored{No Output} & N/A & \cored{No Output} \\
				\hline
				\multirow{5}{*}{Eg6} & Ours & \multirow{5}{*}{1.00} & $(0.97,0.00)\pm(0.00,0.00)$ & 1 & $15.86|63.35$ \\
				& \simpleIVPdirect\ & & \cored{Timeout} & N/A & \cored{Timeout} \\
				& VNODE-LP & & $(0.97,0.00)\pm(0.00,0.00)$ & 1.00 & 11.06 \\
				& CAPD & & $(0.97,0.00)\pm(0.00,0.00)$ & 1.00 & 16.39 \\
				\hline
				\multirow{10}{*}{Eg7} & Ours & \multirow{5}{*}{1.00} & $(-6.9,3.0,35.1)\pm(0.03,0.01,0.04)$ & 1 & $0.12|0.65$ \\
				& \simpleIVPdirect\ & & $(-6.9,3.0,35.1)\pm(37.40,228.27,225.31)$ & $5692.51$ & 7.35 \\
				& VNODE-LP & & $(-6.95,3.00,35.14)\pm(0.03,0.01,0.04)$ & 1.03 & 0.66 \\
				& CAPD & & $(-6.95,3.00,35.14)\pm(0.03,0.01,0.04)$ & 1.02 & 0.10 \\
				\cline{2-6}
				& Ours & \multirow{5}{*}{4.00} & $(-4.75,-0.01,29.07)\pm(0.09,0.09,0.11)$ & 1 & $4.03|12.55$ \\
				& \simpleIVPdirect\ & & \cored{Timeout} & N/A & \cored{Timeout} \\
				& VNODE-LP & &
				\cored{Invalid}
				& N/A & \cored{Invalid} \\
				& CAPD & & \cored{No Output} & N/A & \cored{No Output} \\
				\hline
				\multirow{10}{*}{Eg8} & Ours & \multirow{5}{*}{1.00} &
				$(-1.73,1.87,0.03)\pm(0.15,0.17,0.00)$ & 1 & $0.03|0.16$
				\\
				& \simpleIVPdirect\ & &
				$(-1.73,1.87,0.03)\pm(0.27,0.29,0.00)$ & 1.72 & 0.01 \\
				& VNODE-LP & & $(-1.74,1.86,0.03)\pm(0.15,0.18,0.00)$ & 1.04 & 0.08 \\
				& CAPD & & $(-1.73,1.87,0.03)\pm(0.15,0.17,0.00)$ & 1.02 & 0.02 \\
				\cline{2-6}
				& Ours & \multirow{5}{*}{4.00} & $(1.95,-2.89,0.05)\pm(0.20,0.23,0.00)$ & 1 & $0.08|0.45$ \\
				& \simpleIVPdirect\ & & $(1.95,-2.89,0.05)\pm(11.67,8.86,67.65)$ & 292.99 & 1.45 \\
				& VNODE-LP & & $(1.96,-2.88,0.05)\pm(0.21,0.24,0.00)$ & 1.04 & 0.28 \\
				& CAPD & & $(1.95,-2.89,0.05)\pm(0.21,0.23,0.00)$ & 1.02 & 0.07 \\
				\hline
			\end{tabular}
		}
		\caption{Performance comparison of \EndCover\ with existing
		methods under
			$\veps = 1.0$.
			For our method, the reported time consists of two values separated by
			$a \mid b$, where $a$ denotes the computation time using the End cover
			and $b$ denotes the computation time using the Boundary cover.
			The reported $B_1$ for our method is obtained by taking the minimal
			enclosing box of the Boundary cover.}
		
		\label{tab:main}
	\end{table*}
}
\issacArxiv[\begin{table}
	\centering
	\scriptsize
	\setlength{\tabcolsep}{3pt}
	\begin{tabular}{@{}c|l|c|c|c|c@{}}
		\toprule
		\textbf{Eg} & \textbf{Method} & $T$ & $B_1 = B_{cen}(rad)$ & $\frac{\wmax(\text{other})}{\wmax(\text{ours})}$ & \textbf{Time (s)} \\
		\midrule
		\multirow{12}{*}{Eg1} & Ours & 2.00 & $B_{(0.08,0.57)}(0.02,0.02)$ & 1 & $0.02|0.05$ \\
		& \simpleIVPdirect & & $B_{(0.08,0.58)}(0.16,0.13)$ & 6.66 & 0.52 \\
		& VNODE-LP & & $B_{(0.08,0.58)}(0.04,0.05)$ & 2.26 & 0.17 \\
		& CAPD & & $B_{(0.08,0.58)}(0.05,0.05)$ & 2.12 & 0.01 \\
		\cline{2-6}
		& Ours & 4.00 & $B_{(1.46,0.19)}(0.40,0.05)$ & 1 & $0.13|0.08$ \\
		& \simpleIVPdirect & & \cored{Timeout} & N/A & \cored{Timeout} \\
		& VNODE-LP & & $B_{(1.43,0.34)}(3.23,0.91)$ & 8.07 & 0.39 \\
		& CAPD & & $B_{(1.45,0.19)}(3.38,0.88)$ & 8.35 & 0.02 \\
		\cline{2-6}
		& Ours & 5.50 & $B_{(1.14,2.95)}(0.67,0.49)$ & 1 & $3.42|1.21$ \\
		& \simpleIVPdirect & & \cored{Timeout} & N/A & \cored{Timeout} \\
		& VNODE-LP & & \cored{Invalid} & N/A & \cored{Invalid} \\
		& CAPD & & \cored{No Output} & N/A & \cored{No Output} \\
		\midrule
		\multirow{8}{*}{Eg2} & Ours & 1.00 & $B_{(-2.13,0.57)}(0.25,0.21)$ & 1 & $0.02|0.04$ \\
		& \simpleIVPdirect & & $B_{(-2.13,0.57)}(0.26,0.23)$ & 1.03 & 0.51 \\
		& VNODE-LP & & $B_{(-2.15,0.58)}(0.90,3.49)$ & 13.95 & 0.17 \\
		& CAPD & & $B_{(-2.13,0.56)}(0.29,0.28)$ & 1.15 & 0.02 \\
		\cline{2-6}
		& Ours & 2.00 & $B_{(-1.41,0.96)}(0.38,0.35)$ & 1 & $0.77|0.14$ \\
		& \simpleIVPdirect & & \cored{Timeout} & N/A & \cored{Timeout} \\
		& VNODE-LP & & \cored{Invalid} & N/A & \cored{Invalid} \\
		& CAPD & & \cored{No Output} & N/A & \cored{No Output} \\
		\midrule
		\multirow{4}{*}{Eg3} & Ours & 1.00 & $B_{(-0.60,-6.69)}(0.00,0.18)$ & 1 & $0.02|0.07$ \\
		& \simpleIVPdirect & & $B_{(-0.60,-6.69)}(0.01,0.19)$ & 1.01 & 4.11 \\
		& VNODE-LP & & $B_{(-0.60,-6.69)}(0.00,0.19)$ & 1.05 & 0.20 \\
		& CAPD & & $B_{(-0.60,-6.69)}(0.01,0.19)$ & 1.01 & 0.02 \\
		\midrule
		\multirow{8}{*}{Eg4} & Ours & 1.00 & $B_{(0.28,-0.58)}(0.15,0.16)$ & 1 & $0.01|0.02$ \\
		& \simpleIVPdirect & & $B_{(0.28,-0.59)}(0.16,0.17)$ & 1.06 & 0.01 \\
		& VNODE-LP & & $B_{(0.28,-0.59)}(0.16,0.18)$ & 1.10 & 0.03 \\
		& CAPD & & $B_{(0.28,-0.59)}(0.16,0.17)$ & 1.06 & 0.01 \\
		\cline{2-6}
		& Ours & 4.00 & $B_{(-0.65,0.25)}(0.24,0.57)$ & 1 & $0.12|0.06$ \\
		& \simpleIVPdirect & & $B_{(-0.68,0.34)}(1.01,2.12)$ & 3.71 & 0.01 \\
		& VNODE-LP & & $B_{(0.59,2.53)}(22.57,88.77)$ & 155.74 & 0.09 \\
		& CAPD & & $B_{(-0.68,0.34)}(7.51,25.53)$ & 44.79 & 0.01 \\
		\midrule
		\multirow{8}{*}{Eg5} & Ours & 1.00 & $B_{(1.76,0.17)}(0.11,0.10)$ & 1 & $0.03|0.07$ \\
		& \simpleIVPdirect & & $B_{(1.76,0.17)}(0.17,0.10)$ & 1.55 & 0.01 \\
		& VNODE-LP & & $B_{(1.76,0.17)}(0.10,0.10)$ & \coblue{0.95} & 0.09 \\
		& CAPD & & $B_{(1.76,0.17)}(0.18,0.10)$ & 1.66 & 0.02 \\
		\cline{2-6}
		& Ours & 4.00 & $B_{(1.68,0.68)}(0.08,0.10)$ & 1 & $1.22|0.13$ \\
		& \simpleIVPdirect & & \cored{Timeout} & N/A & \cored{Timeout} \\
		& VNODE-LP & & $B_{(1.68,0.68)}(0.07,0.11)$ & 1.13 & 0.41 \\
		& CAPD & & \cored{No Output} & N/A & \cored{No Output} \\
		\midrule
		\multirow{4}{*}{Eg6} & Ours & 1.00 & $B_{(0.97,0.00)}(0.00,0.00)$ & 1 & $15.86|63.35$ \\
		& \simpleIVPdirect & & \cored{Timeout} & N/A & \cored{Timeout} \\
		& VNODE-LP & & $B_{(0.97,0.00)}(0.00,0.00)$ & 1.00 & 11.06 \\
		& CAPD & & $B_{(0.97,0.00)}(0.00,0.00)$ & 1.00 & 16.39 \\
		\midrule
		\multirow{8}{*}{Eg7} & Ours & 1.00 & $B_{(-6.9,3.0,35.1)}(0.03,0.01,0.04)$ & 1 & $0.12|0.65$ \\
		& \simpleIVPdirect & & $B_{(-6.9,3.0,35.1)}(37.4,228.2,225.3)$ & 5692.51 & 7.35 \\
		& VNODE-LP & & $B_{(-6.9,3.0,35.1)}(0.03,0.01,0.04)$ & 1.03 & 0.66 \\
		& CAPD & & $B_{(-6.9,3.0,35.1)}(0.03,0.01,0.04)$ & 1.02 & 0.10 \\
		\cline{2-6}
		& Ours & 4.00 & $B_{(-4.75,-0.01,29.07)}(0.09,0.09,0.11)$ & 1 & $4.03|12.55$ \\
		& \simpleIVPdirect & & \cored{Timeout} & N/A & \cored{Timeout} \\
		& VNODE-LP & & \cored{Invalid} & N/A & \cored{Invalid} \\
		& CAPD & & \cored{No Output} & N/A & \cored{No Output} \\
		\midrule
		\multirow{8}{*}{Eg8} & Ours & 1.00 & $B_{(-1.73,1.87,0.03)}(0.15,0.17,0.00)$ & 1 & $0.03|0.16$ \\
		& \simpleIVPdirect & & $B_{(-1.73,1.87,0.03)}(0.27,0.29,0.00)$ & 1.72 & 0.01 \\
		& VNODE-LP & & $B_{(-1.74,1.86,0.03)}(0.15,0.18,0.00)$ & 1.04 & 0.08 \\
		& CAPD & & $B_{(-1.73,1.87,0.03)}(0.15,0.17,0.00)$ & 1.02 & 0.02 \\
		\cline{2-6}
		& Ours & 4.00 & $B_{(1.95,-2.89,0.05)}(0.20,0.23,0.00)$ & 1 & $0.08|0.45$ \\
		& \simpleIVPdirect & & $B_{(1.95,-2.89,0.05)}(11.67,8.86,67.65)$ & 292.99 & 1.45 \\
		& VNODE-LP & & $B_{(1.96,-2.88,0.05)}(0.21,0.24,0.00)$ & 1.04 & 0.28 \\
		& CAPD & & $B_{(1.95,-2.89,0.05)}(0.21,0.23,0.00)$ & 1.02 & 0.07 \\
		\bottomrule
	\end{tabular}
	\caption{\textbf{Performance comparison of EndCover with existing methods ($\veps=1.0$).} For our method, time is shown as $\text{End cover}|\text{Boundary cover}$. $B_1$ is shown as $B_{cen}(rad)$ where $cen$ is center and $rad$ is radius vector.}
	\label{tab:main}
\end{table}
]{}

	\refTab{main} contains the performances of
	four validated interval IVP solvers on the models 
	in \refTab{problems}.  These solvers are:
	\\(1) “Ours” refers to the \EndCoverAlgo\ in this paper.
	\\(2) \simpleIVPdirect\ (see \cite{zhang-yap:ivp:25arxiv}),
	viewed as a ``baseline algorithm'' since it uses the
	standard iterative scheme that alternates between stepA and
	stepB, with stepB using the direct method.
	\\(3) “VNODE-LP” from Nedialkov
		\cite{vnode:home,nedialkov+2:validated-ode:99}.
		It implements Lohner's method.
	\\(4) “CAPD” \cite{wilczak-zgliczynski:lohner:11}
		that implements Cr-Lohner method (using $r=3$ in
		their software).

	Each solver is tasked with
	computing the end-enclosure for $\ivp(B_0, T)$
	under identical conditions. Since “Ours” outputs
	a set of boxes, we take the smallest bounding box of their union.
	Thus, the reported enclosure is a
	over-approximation of our true output.
	
	\dt{Efficacy of our method.}
	Define the \dt{efficacy ratio} of some method “Other” 
	relative to “Ours” as
			$\rho(Others)\as \frac{\wmax(Others)}{\wmax(Ours)}$.
	So $\rho(Other)>1$ means “Ours” is more efficacious.
	The efficacy column in Table~\ref{tab:main} demonstrates
	that $\rho(Other)>1$ for all but one entry ($0.95$ in blue).
	Note that $\rho(Other)$ increases as $T$ increases.

	\dt{Robustness of our method.}
	\refTab{main} shows 3 kinds of abnormal behavior:
	\cored{Timeout} (if it takes longer than 30 minutes),
	\cored{No Output} (the code stops without output, or
	returns error),
	and
	\cored{Invalid}.
	The \cored{No Output} behavior is seen with \capdCr: see
			Eg1 with $T=5.5$,
			Eg2 with $T=2$,
			Eg5 with $T=4$, and
			Eg7 with $T=4$.
	The \cored{Invalid} behavior is seen in \vnodelp, 
	meaning that its step size has gone below some threshhold: see
			Eg1 with $T=5.5$,
			Eg2 with $T=2$, and
			Eg7 with $T=4$.
	However, “Ours” do not show any of these abnormal behavior,
	i.e., it is robust.
	
	\ignore{
	It is worth noting that VNODE-LP and CAPD can compute enclosures
	efficiently for small integration times $T$. However, as $T$
	increases these enclosures tend to grow rapidly; consequently some
	computations—most notably the $QR$ decompositions used in Lohner-type
	procedures—can become numerically unstable and the method may fail to
	produce results (see Eg1, Eg2, Eg4). It is unclear whether more
	stable version of CAPD exists. However, the experimental data for
	Example 1 at ($T=2,4,5.5$) indicate that the ratio
	$\wmax(\mathrm{CAPD})/\wmax(\mathrm{ours})$ at $T=5.5$ if a result
	can be produced, is expected to be larger than its value at $T=4$.
	}%

	\ignore{ 
	\dt{Mitigation of the Wrapping Effect:}
	standard iterative IVP methods suffers from the
	wrapping effect, i.e., the
	end enclosures inflates over time, forcing the
	step size to decrease rapidly. This behavior significantly increases
	the computational cost and, in many cases, prevents the computation
	from completing.  Our \EndCover\ and \BoundaryCover\
	suffers less from this.  
	}%
	
\ssect{Comparing \EndCover\ with \BoundaryCov}
	The standard iterative IVP algorithms (e.g., \simpleIVPdirect) 
	suffers from the wrapping effect, and must use Lohner's
	matrix transformation.  However, our \EndCover\
	and \BoundaryCov\ reduces the wrapping effect by tracking
	the boundary with smaller boxes.
	This leads to improved numerical stability.
	Furthermore, our \Refine\ algorithm
	prevents the intermediate enclosures from becoming
	excessively large; thus we avoid unnecessary step-size
	reductions leading to loss of precision.
	When $T$ is small, the \BoundaryCov\ method is slower than
	the \EndCover\ method. 
	But when $T$ is large, the Boundary method becomes faster
	since it only needs to handle the boundary; see Eg1, Eg2, Eg4, Eg5 in \refTab{main}.


%% file: inc/6.tex
\sect{Conclusion}
	We have designed a complete validated
	algorithm for the reachability form of the IVP problem, called
	the End Cover Problem.  We also offer
	an alternative technique based on covering the boundary
	of the End Cover as a lower-dimensional problem
	(and thus possibly faster).
	Preliminary implementations show that our algorithms
	are practical, robust, and outperforms state-of-the-art
	validated algorithms (\capdCr\ and \vnodelp).

%% file: inc/appendixA.tex
\newpage
\appendix
\section{Appendix: Proofs} 
	The numberings of lemmas and theorems
	in this Appendix are the same as the corresponding
	results in the text; they are also hyperlinked to the
	text.

	\ignore{
	\bthmDIY[Theorem \ref{thm:bundle}]{
			({\sc Existence of $\bff$-bundles})
		{\em  \ \\
		If $\bff:\RR^n\to\RR^n$ is locally Lipschitz, and
		$B\ib \valid_\bff(H)$, then
		the $\bff$-bundle $\Phi_\bff(B,H)$ exists.
		}
	}
	\chee{THIS IS NO GOOD}
	\bpf
	For each $\bfp\in B$, we have an $\bff$-bundle
	$\Phi_\bff(Ball_\bfp(r),H)$ for some $r=r(\bfp)>0$ by \refthm{pl}.
	The union of all these $\bff$-bundles is $\ol\Phi$,
	the $\bff$-bundle on $(\olB,H)$ where $\olB$ is
	the union of $Ball_\bfp(r(\bfp))$ for $\bfp\in B$.
	Since $B\ib \olB$, the desired bundle $\Phi_\bff(B,H)$
	is just the restriction $\ol\Phi$ to $[0,H]\times B$.
	\epf}
		\issacArxiv{
	\bthmDIY[Theorem \ref{thm:homeo}]{
		{\em  \ \\
		Let $\Phi:[0,h]\times B\to\RR^n$ be a trajectory
		bundle where $B\in\intbox\RR^n$.
		Then the end map $\End_\Phi$ of \refeq{endmap}
		is a homeomorphism from $B$ onto the end-slice $\Phi|^h$.
		}
	}
		\bpf
	The map \refeq{endmap} has two properties:
	\benum[(a)]
	\item 
	It is continuous
	(this follows from the continuity of $\Phi$).
	\item 
	It is injective
	(this is the stratification property).
	\eenum
	If $U$ is the interior of $B$, then Brouwer's
	Invariance of Domain implies $U$ is homeomorphic
	to $\End_\Phi(U)$. 
	But the closure of $\End_\Phi(U)$ is the end-slice $\Phi|^h$.
	Part(b) above says that
	$\End_\Phi: B\to \Phi|^h$ is injective.
	This implies that $\End_\Phi$ is a homeomorphism
	(\cite[Theorem 4.17, p.90]{rudin:principles:bk}).
	\epf}

\ignore{
\blemDIY[Lemma \ref{lem:valid}]{
		{\em  \ \\
		The set $\valid_\bff(H)$ is an open set for any $H>0$.
		}
	}
	\bpf
		There is nothing to prove if $\valid(H)$ is empty
		(since an empty set is open).
		Otherwise, let $\bfp^*\in \valid(H)=\valid_\bff(H)$ 
		It suffices to prove that $\valid(H)$
		contains an open neighborhood $U^*$ of $\bfp^*$.
		Since $\bfp^*\in\valid(H)$, 
		there is a $\bff$-trajectory $\phi^*$
		with time span $[0,H]$.
		For each $t\in [0,H]$, 
			the	Local Picard-Lindel\"of Bundle Theorem
		implies that there exists a $\bff$-bundle $\Phi_t$ on
			$[0,\veps(t)]\times B(\phi^*(t),r(t))$.
		Since $\bff$ is autonomous, we may shift the time-span so
		that $\Phi_t$ is a $\bff$-bundle on the cylinder
			$$\cyl(t) =[t,t+\veps(t)]\times Ball_{\phi^*(t)}(r(t)).$$
		Thus the set $\set{(t,t+\veps(t)): t\in [0,h]}$
		is an open cover of the interval $[\half\veps(0),h]$.
		By compactness of $[\half \veps(0), h]$, 
		there is a finite set
			$$C=\set{(t_i,t_i+\veps(t_i)): i=0\dd m}$$
		that forms a cover of $[\half\veps(0),h]$.
		Without loss of generality, we may assume that
			$$0= t_0< t_1 <\cdots t_m < h$$
		and the set $C$ is a minimal cover,
		i.e., if we omit any $(t_i, t_i+\veps(t_i))$
		from $C$, it would not cover $[\half\veps(0),h]$.
		For each $i=1\dd m-1$, choose 
			$s_i\in (t_{i-1},t_{i-1}+\veps(t_{i-1}))
				\cap (t_{i},t_{i}+\veps(t_{i}))$.
		Minimality of $C$ implies that
			$0<s_1<s_2<\cdots < s_{m-1}<h$
		Also set $s_0=0$ and $s_m=h$. 
		
		To prove our theorem, we show how to use the
		trajectory $\phi^*$ to construct the open neighhood $U^*$:
		Recall that $\Phi_{t_i}$ ($i=0\dd m-1$) is a $\bff$-bundle
		  $\cyl(t_i)=
		  	[t_i,t_i+\veps(t_i)]\times Ball_{\phi^*(t_i)}(r(t_i))$.
		But since $[s_i,s_{i+1}]\ib [t_i,t_i+\veps(t_i)]$, we
		can define $\Phi_{(i)}$ to be the restriction of
		$\Phi_{t_i}$ to the domain $[s_i,s_{i+1}]\times B_i$ where
		$B_i$ is the $s_i$-slice of $\Phi_{t_i}$.
		We now have a sequence of $\bff$-bundles
			$$\Phi^{(0)}, \Phi^{(1)}\dd \Phi^{(m-1)}.$$
		Note that the trajectory $\phi^*$ when restricted
		to the time-span $[s_i,s_{i+1}]$ is actually
		a trajectory of the bundle $\Phi^{(i)}$.
		Let $B'_{i}$ denote the end-slice of $\Phi^{(i)}$.
		Then map $E_i = \End_{\Phi^{(i)}}$
		gives us a homeomorphism from $B_i$ to $B'_{i}$.
		Moreover, $B'_{i}\cap B_{i+1}$ is non-empty since
		the trajectory $\phi^*$ passes through them both
		at time $s_{i+1}$:  $\phi^*(s_{i+1}\in B_{i+1}\cap B'_{i}$.

		We want to construct a ``composition'' of the $E_i$'s,
		analogous to $E_1\circ E_2\circ \cdots \circ  E_m$.
		First, let us define $E_{i-1}*E_i$
		(analogous to $E_{i-1}\circ E_i$):
		from $E_{i-1}: B_{i-1}\to B'_{i-1}$
		and $E_{i}: B_{i}\to B'_{i}$, we get the homeomorphism
			$$E_{i-1}*E_{i}:
				E_{i-1}\inv( B'_{i-1}\cap B_i)
				\to
				E_{i}( B'_{i-1}\cap B_i)$$
		where we may define 
				$B_{i-1,i}\as E_{i-1}\inv( B'_{i-1}\cap B_i)$
				and
				$B'_{i-1,i}\as E_{i}( B'_{i-1}\cap B_i)$.
		Inductively, for $i=m-1, m-2\dd 1$, define 
			$$(E_i*\cdots * E_m): B_{i,m}\to B'_{i,m}$$
		where
			$$B_{i,m}=E\inv_{i-1}(B'_{i}\cap B_{i+1,m}),
			\qquad
			B'_{i,m}=E_{i}(B'_{i}\cap B_{i+1,m})$$
		Thus, the final composition we want is the homeomorphism
			$E_1*\cdots *E_m: B_{1,m}\to B'_{1,m}$.
		where $B_{1,m}, B'_{1,m}$ are all open sets
	\epf}


%
\issacArxiv[Proof of Lemma \ref{lem:Numberofstages}]{
	\blemDIY[Lemma \ref{lem:Numberofstages}]{
		{\em  \ \\
		The number of stages in
		\endEnc$(E_0,\bfveps; \bfp,H)$ is at most
		$1+\floor{H/\olh}$ where
		$
		\olh=\olh(\olE,H,\bfveps)$, see \refeq{olh}
		and
		$
		\olE =\image(\ivp(E_0,H))+Box(\bfveps).
		$
		}
	}}
		\bpf
	Since $\ivp(B_0,H)$ is valid, $\olE$ is bounded. The set
	$\olB=\olB(\olE,H,\bfveps)$ is also bounded as it is constructed from
	bounded sets using continuous
	operations. Since $\bff^{[k]}$ is continuous on the compact set
	$\olB$, each $\olM_i$ ($i=1\dd n$)
	in the definition \refeq{M_i} of
	$\olh=\olh(\olE,H,\bfveps)$ is finite.  Thus,
	$
	0<\olh \le
	\min_{i=1}^n \Big(\veps_i/\olM_i\Big)^{1/k} .  $
	Supppose the algorithm terminates after $m$ stages.
	Then the time sequence of the final scaffold
	has the form $\bft=(t_0,t_1\dd t_m)$ where $(t_0,t_m)=(0,H)$.
	The $(j-1)$th stage was extended to the $j$th
	stage by calling $\stepA(E_{j-1},H-t_{j-1},\bfveps)\to (h_j,F_j)$.
	Clearly, $h_j= t_j-t_{j-1}$. 
	By \refeq{olh}, 
	$$h_j \ge 			\min\set{H-t_{j-1},\min_{i=1}^n
		\Big(\tfrac{\veps_i}{M_i}\Big)^{1/k} }.$$
	If \(h_j = H - t_{j-1}\), then \(j\) must be the final stage (\(j=m\)).
	Therefore, for every \(j<m\),
	$
	h_j \;\ge\; \min_{i=1}^n \Bigl(\tfrac{\veps_i}{M_i}\Bigr)^{1/k}.
	$
	We claim that for all \(j = 1,\dots,m-1\), one has \(h_j \ge \olh\).
	This immediately yields
	$
	(m-1)\,\olh \le H,
	\,\text{hence}\,
	m \le 1 + \lfloor H/\olh \rfloor .
	$
	To prove the claim, fix any \(j<m\). Then
	\beqarrys
	h_j &\ge& \min_{i=1}^n
	\Big(\veps_i/M_i\Big)^{1/k} 
	& \text{(since $j<m$)}\\
	&\ge& \min_{i=1}^n
	\Big(\veps_i/\olM_i\Big)^{1/k} 
	& \text{(since $M_i \le \olM_i$)}\\
	&\ge& \olh
	& \text{(by definition of $\olh$)}\\
	\eeqarrys
	
	The above inequality $M_i \le \olM_i$ comes from the 
	inclusion $$\olB(E_{j-1},H-t_{j-1},\bfveps) \ib \olB(\olE,H,\bfveps),$$ and the definition of 
	$M_i,\olM_i$ in terms of these two sets.
	This establishes \(m-1 \le H/\olh\) as required.
	\epf

\issacArxiv[Proof of Lemma \ref{lem:complexityExtend}]{
	\blemDIY[Lemma \ref{lem:complexityExtend}]{{\sc (Extend)}
	{\em  \ \\
	Given a scaffold \(\stage\) with \(m\) stages and parameters
	\(\bfveps>0\) and \(H>0\), the subroutine \stepA\ executed within
	\(\stage.\Extend(\bfveps,H)\) performs at most
	$$
	\left\lceil \log_2\!\left(\frac{H}{\olh}\right) \right\rceil
	$$
	iterations, where
	\(\olh=\olh(E_m(\stage),H,\bfveps)\) is defined in \refeq{olh}
	}
}}
	\bpf
	The \Extend\ subroutine will call 
	\stepA\, which
	performs a binary search by halving the 
	step size $H$ until an admissible pair $(h>H/2, F)$ is found. 
	The number of halvings required is $O(\log_2(H / \olh))$, 
	as $H$ is reduced to a minimum step size bounded 
	below by $\olh > 0$.
	\epf
	
	\ignore{
	\blemDIY[Lemma \ref{lem:terminationRefine}]{
		{\em  \ \\
				The modified \Refine\ subroutine terminates.
		}
	}
		\bpf
	Suppose, to the contrary, that the modified \Refine\ never terminates.  
	Then it produces infinitely many phases and an infinite sequence of chains
	\[
	C_1 < C_2 < C_3 < \cdots,
	\]
	where we write 
	\(
	C_p = (k(1), k(2) \dd k(m)) < C_q = (k'(1), k'(2) \dd k'(m))
	\)
	if \(k(i) < k'(i)\) for every \(i = 1 \dd m\).
	
	In every phase,  \(\wmax(E_0(\stage))\) is halved whenever
	Condition  1 \refeq{con-1} fails, and therefore it must eventually satisfy Condition 1 \refeq{con-1}.
	Likewise, for each chain \(C_i\), the value \(\Delta(C_i)\) is reduced by at least a factor of~2
	and hence Condition 2 \refeq{con-2} must eventually be met
	for all stages as well.
	
	Once both Condition~1 and for all stages Condition~2 hold
	simultaneously,  
	Condition~3 becomes active, causing \Refine\ to halt.
	This contradicts the assumption of nontermination.
	\epf}
	
	\issacArxiv[Proof of Lemma \ref{lem:complexityRefine}]{
		\blemDIY[Lemma \ref{lem:complexityRefine}]{{\sc (Refine)}
		{\em  \ \\
			Let $\stage$ be an $m$-stage scaffold, and $\veps_0>0$.  
			For each stage $i$ write
			$\Delta t_i = t_{i+1}(\stage)-t_i(\stage)$ and
			\[
			G_i(\stage) = ((\pi_i, \bfg_i),
			(\ell_i, \bfE_i, \bfF_i),
			(\ol\bfmu^1_i, \ol\bfmu^2_i), 
			(\delta_i, \wh{h}_i)).
			\]
			The number of phases in $\stage.\Refine(\veps_0)$ is at most
			$1+N_0$ where
			\bitem
			\item $N_0 \as \max\set{N_1,N_2}$,
			\item $N_1 \as \Bigl\lceil \log_2
			\frac{2 \, \wmax(E_0(\stage)) \, e^{\olmu}}{\veps_0}
			\Bigr\rceil$,
			\item $N_2 \as \#\text{EulerTube} + \#\text{Bisection}$
			\item $\#\text{EulerTube} \as \Bigl\lceil \log_2
			\frac{8 m \, \Delta_{\max}
				\, e^{\olmu}}{\veps_0} \Bigr\rceil$
			\item $\#\text{Bisection} \as 
			\Biggl\lceil 
			\displaystyle{\max_{i=1,\dots,m, p\in I_i}} \left(
			\log_2 \Biggl(
			\frac{\Delta t_i}{\, h^{\euler} \Bigl( \Delta t_i, M_i, 
				p, 
				\delta_i'
				\Bigr) }
			\Biggr)\right) \Biggr\rceil$\\
			\item Moreover,
			$I_i= \big[
			\min_{\bfq\in \pi(F_i(\stage))}
			(\mu_2(J_{\bfg_i}(\bfq))),
			\displaystyle\max_{i=1\dd m}(max \ol\bfmu_i^2)\big]$,
			\item 
			$ \olmu \as \max_{i=1\dd m}(max \ol\bfmu_i^1)$,
			\item
			$\Delta_{\max}  \as  \max_{i=1,\dots,m} \delta_i$,
			\item 
			$M_i \as \|\bfg_i^{[2]}(\pi(F_i(\stage)))\|_2$,
			\item $\delta_i'\as \TransformBound
			\left(\min\!  \left( \delta_i,
			\frac{\veps_0}{16 m e^{\olmu}}\right),
			\pi,F_i(\stage)\right).$
			\eitem
		}
	}}
	\bpf
	$N_1$ is the number of phases needed to satisfy condition \refeq{con-1}.
	Since in each phase we halve $\wmax(E_0(\stage))$, after
	$$
	N_1 \as \Bigl\lceil \log_2 \frac{2 \, \wmax(E_0(\stage)) \, e^{\olmu}}{\veps_0} \Bigr\rceil
	$$
	phases, condition~\refeq{con-1} is satisfied.
	
	Next, we consider the phases required for condition~\refeq{con-2}.
	Fix a stage $i$.
	Note that in each phase, the value $\delta_i$ is either halved or left unchanged; once $\delta_i$ satisfies~\refeq{con-2} it is never modified again.
	
	By condition~\refeq{con-2},
	$
	\delta_i < \frac{\veps_0}{8 m e^{\olmu}}$ and
	$
	\delta_i \le \Delta_{\max},
	$
	so the number of phases in which $\delta_i$ is halved is at most
	\[
	\#\text{EulerTube} \as \Bigl\lceil \log_2 \frac{8 m \, \Delta_{\max} \, e^{\olmu}}{\veps_0} \Bigr\rceil.
	\]
	
	Now we count the phases where $\delta_i$ stays unchanged.
	During $\Refine$, if $\delta_i$ is not changed in a phase, then \Bisect is called in that phase.
	Hence it suffices to count the number of calls to \Bisect.
	A call to \Bisect occurs exactly when the current mini‑step size, i.e., $\dfrac{\Delta t_i}{2^{\ell}}$, exceeds $h^{\euler}$.
	Thus we need a lower bound for $h^{\euler}$.
	
	Recall that whenever the logarithmic norm $\mu$ is 
	positive, we transform the system by $\pi$, which 
	yields a new system with a negative logarithmic norm. 
	Consequently 
	(see~\cite[Eq.~(12)]{zhang-yap:ivp:25arxiv})
	\[
	h^{\euler}(H,\olM,\mu,\delta)=
	\min\!\Bigl\{H,\,
	\frac{2\mu\delta}{M(e^{\mu H}-1)-\mu^2\delta}\Bigr\}.
	\]
	 $h^{\euler}(H,\olM,\mu,\delta)$ is decreasing in $\olM$ and increasing in $\delta$.  
	During refinement the set $F_i(\stage)$ shrinks, hence
	$
	M_i = \|\bfg_i^{[2]}(\pi(F_i(\stage)))\|_2
	$
	provides an upper bound for the parameter $\olM$ in stage~$i$.
	Moreover, condition~\refeq{con-2} guarantees that 
	$\min\!\bigl(\delta_i,\; \frac{\veps_0}{16 m e^{\olmu}}\bigr)$ is a lower bound for $\delta(G_i(\stage))$.
	Therefore $\delta_i'$ is a lower bound for the $\delta$‑parameter in stage~$i$.  
	Finally, by definition of $I_i$, the parameter $\mu$ in the $i$‑th stage always lies in $I_i$.
	
	Consequently, for stage~$i$ the Euler step size is bounded below by
	$\displaystyle\min_{p\in I_i} h^{\euler}\!\Bigl(\Delta t_i,\; M_i,\; p,\; \delta_i'\Bigr)$.
	Hence the number of calls to \Bisect\ for the $i$‑th stage is at most
	$
	\max_{p\in I_i} \,
	\log_2 \Biggl(
	\frac{\Delta t_i}{\, h^{\euler} \bigl( \Delta t_i, M_i, p, \delta_i' \bigr) }
	\Biggr).
	$
	
	Thus, $N_2$ is the number of phases needed to satisfy
	condition \refeq{con-2} for every stage.
	\epf
	
	\issacArxiv[Proof of Theorem \ref{thm:complexityEndCover}]{
		\bthmDIY[Theorem \ref{thm:complexityEndCover}]{
		{\em  \ \\
				Let $\olB = \image(\ivp(B_0,H)) + Box(\veps)$ and
			$\olmu = \mu_2(J(\bff(\olB)))$.
			Then $\EndCover(B_0,H,\veps)$ will call \endEnc\ at most
			\[
			\Theta\!\left(
			\left(\frac{2\wmax(B_0)e^{\olmu}}{\veps}\right)^n
			\right)
			\]
			times.
		}
	}}
		\bpf  
	Each call of \endEnc\ returns two boxes $\ulB_i$ and $\olB_i$.  
	We bound the number of indices $i$ for which the collection $\{\ulB_i\}$ can cover $B_0$.
	
	From \refeq{con-1}, every box $\ulB_i$ satisfies a uniform positive bound on its size:
	$\frac{\veps}{4 e^{\olmu H}}<\wmax(\ulB)<\frac{\veps}{2 e^{\olmu H}}$. 
	Consequently, starting from a initial box $B$, the number of subdivisions needed to obtain a box whose size is below this lower bound is at most  
	$
	\Bigl\lceil \log_2 \frac{4\,\wmax(B)\,e^{\olmu H}}{\veps} \Bigr\rceil .
	$
	Hence, along each coordinate direction, the number of such subdivisions is bounded by the above expression.  
	In $n$ dimensions, it follows that at most  
	\[
	O\!\left(
	\Bigl(\frac{2\,\wmax(B_0)\,e^{\olmu H}}{\veps}\Bigr)^{\!n}
	\right)
	\]
	boxes are created which completes the proof.  
	\epf
	
\ignore{
	\bthm
	Let $\bff:\RR^n\to\RR^n$ be locally Lipschitz and fix $h>0$.  
	Then the set
	\[
	\valid_\bff(h)=\{\bfp\in\RR^n:\ \ivp_\bff(\bfp,h)\ \text{is valid}\}
	\]
	is open.
	\ethm
	
	\bpf
	Let $\bfp_0\in\valid_\bff(h)$. By definition there is a unique
	$\bff$--trajectory
		\[ \bfphi:[0,h]\to\RR^n,\qquad \bfphi(0)=\bfp_0, \]
	solving \refeq{bfx'}.  Set
		\[ K:=\Image_\bff(\{\bfp_0\},h)=\{\bfphi(t):t\in[0,h]\}.  \]
	Since
	$\bfphi$ is continuous on the compact interval $[0,h]$, the set $K$
	is compact.
	
	Because $\bff$ is locally Lipschitz, for every point of $K$ there
	exists a neighborhood on which $\bff$ is Lipschitz. By compactness of
	$K$ we obtain a finite subcover and hence an open set $U\supset K$
	and a constant $L>0$ such that $\bff$ is $L$--Lipschitz on $U$, i.e.
	for all $y_1,y_2\in U$
		\[ \|\bff(y_1)-\bff(y_2)\|\le L\|y_1-y_2\|.  \]
	Since $K\subset U$ and $U$ is open, the distance
		\[ r:=\operatorname{dist}\bigl(K,\RR^n\setminus U\bigr) \]
	is strictly positive.
	
	Choose $\delta>0$ with
		\[ \delta e^{Lh}<r, \]
	for instance
	$\delta=\tfrac{r}{2}e^{-Lh}$. We claim that the open ball
	$B_{\bfp_0}(\delta)$ is contained in $\valid_\bff(h)$.
	
	Fix any $\bfp\in Ball_{\bfp_0}(\delta)$.
	By Picard-Lindel\"of theorems
	there exists a solution $\bfphi_\bfp(t)$ of \refeq{bfx'} with
	$\bfphi_\bfp(0)=\bfp$ defined at least on some nonempty interval
	$[0,\tau)$ with $\tau>0$. Consider the difference
		\[ e(t):=\bfphi_\bfp(t)-\bfphi(t).  \]
	On any time interval where both
	$\bfphi_\bfp(t)$ and $\bfphi(t)$ are defined and remain in $U$ we
	have
		\[ e'(t)=\bff(\bfphi_\bfp(t))-\bff(\bfphi(t)), \]
	and hence by
	the Lipschitz bound
		\[ \|e'(t)\|\le L\|e(t)\|.  \]
	Gronwall's
	inequality then yields, for all $t$ in that common interval,
		\[
		\|e(t)\|\le \|e(0)\|e^{Lt}=\|\bfp-\bfp_0\|e^{Lt}
			\le \|\bfp-\bfp_0\|e^{Lh}<\delta e^{Lh}<r.  \]
	Thus for every $t$ in the
	common interval we have
		\[ \|\bfphi_\bfp(t)-\bfphi(t)\|<r, \]
	which
	implies $\bfphi_\bfp(t)\in U$ (because $\bfphi(t)\in K$ and
	$r=dist(K,\RR^n\setminus U)$). In particular, the solution
	$\bfphi_\bfp(t)$ cannot leave $U$ while it exists.
	
	Now suppose for contradiction that $\bfphi_\bfp$ does not extend to
	the whole interval $[0,h]$. Let $T\le h$ be the right endpoint of the
	maximal interval of existence of $\bfphi_\bfp$. The argument above
	shows that on $[0,T)$ the trajectory $\bfphi_\bfp(t)$ remains in the
	open set $U$ and therefore, since $\bff$ is Lipschitz on $U$,
	Picard--Lindel\"of (or standard extension theory) permits extension of
	$\bfphi_\bfp$ beyond $T$, contradicting maximality of $T$. Hence
	$\bfphi_\bfp$ extends to a solution on the whole $[0,h]$.
	
	Uniqueness on $[0,h]$ follows similarly from the local uniqueness
	given by Picard--Lindelöf together with the fact that every solution
	starting in $B_{\bfp_0}(\delta)$ stays inside $U$ on $[0,h]$; the
	Lipschitz property on $U$ ensures global uniqueness on $[0,h]$.
	Therefore for every $\bfp\in B_{\bfp_0}(\delta)$ the set
	$\ivp_\bff(\bfp,h)$ is a singleton, so
	$B_{\bfp_0}(\delta)\subset\valid_\bff(h)$.
	
	Since $\bfp_0\in\valid_\bff(h)$ was arbitrary, $\valid_\bff(h)$ is
	open. 
	\epf
	}

%% file: inc/appendixB.tex
\newpage
\section{Appendix: More boundary cover Experiments}  
	We provide some additional experimental data.
\ignore{	
	To provide a concrete 
	illustration, we begin with a detailed example that 
	highlights the key features and benefits of our 
	approach in action.
	
	\begin{xample}
		Consider the Volterra system
		
		$$
		\begin{cases}
			x' = 2x - 2xy, \\
			y' = -y + xy,
		\end{cases}
		$$
		
		with initial box $B_0 = [0.9, 1.1] \times [2.9, 3.1]$. We compute
		the end-enclosures of the four edges of $B_0$ at final time $T =
		1$. The four resulting enclosures are:
		
		$$
		\begin{aligned}
			&\text{Edge 1:} && [0.0590893,\ 0.0912428] \times [1.36636,\
			1.49452], \\
			&\text{Edge 2:} && [0.0605758,\ 0.0970096] \times [1.42774,\
			1.56825], \\
			&\text{Edge 3:} && [0.0686275,\ 0.10654] \times [1.38487,\
			1.48567], \\
			&\text{Edge 4:} && [0.0534886,\ 0.083121] \times [1.44681,\
			1.54128].
		\end{aligned}
		$$
		
		The smallest box enclosing all four edge enclosures (i.e., their
		wrapping) is:
		
		$$
		[0.0534886,\ 0.10654] \times [1.36636,\ 1.56825].
		$$
		
		In contrast, a direct enclosure computed for the entire box $B_0$
		(without decomposition) yields:
		
		$$
		[0.0168262,\ 0.137862] \times [1.30876,\ 1.62013].
		$$
		
		This comparison underscores the key strength of 
		our boundary-based propagation strategy. By 
		decomposing the problem into edge-specific 
		evolutions and subsequently combining the 
		results, we achieve a much tighter end-enclosure. 
		Quantitatively, the area of our method's 
		enclosure is approximately 3.52 times smaller 
		than that of the direct approach, illustrating 
		how our technique mitigates the wrapping effect 
		and delivers sharper, more precise bounds that 
		are invaluable for reliable uncertainty 
		quantification in dynamical systems.
		
	\end{xample}
	
}
	The figures below illustrate the trajectories produced by our method
	and demonstrate how the corresponding end enclosures are obtained.
	\begin{figure*}
		\centering
		\begin{minipage}[t]{0.48\linewidth}
			\vspace{0pt}
			\centering
			\includegraphics[height=6.2cm]
			{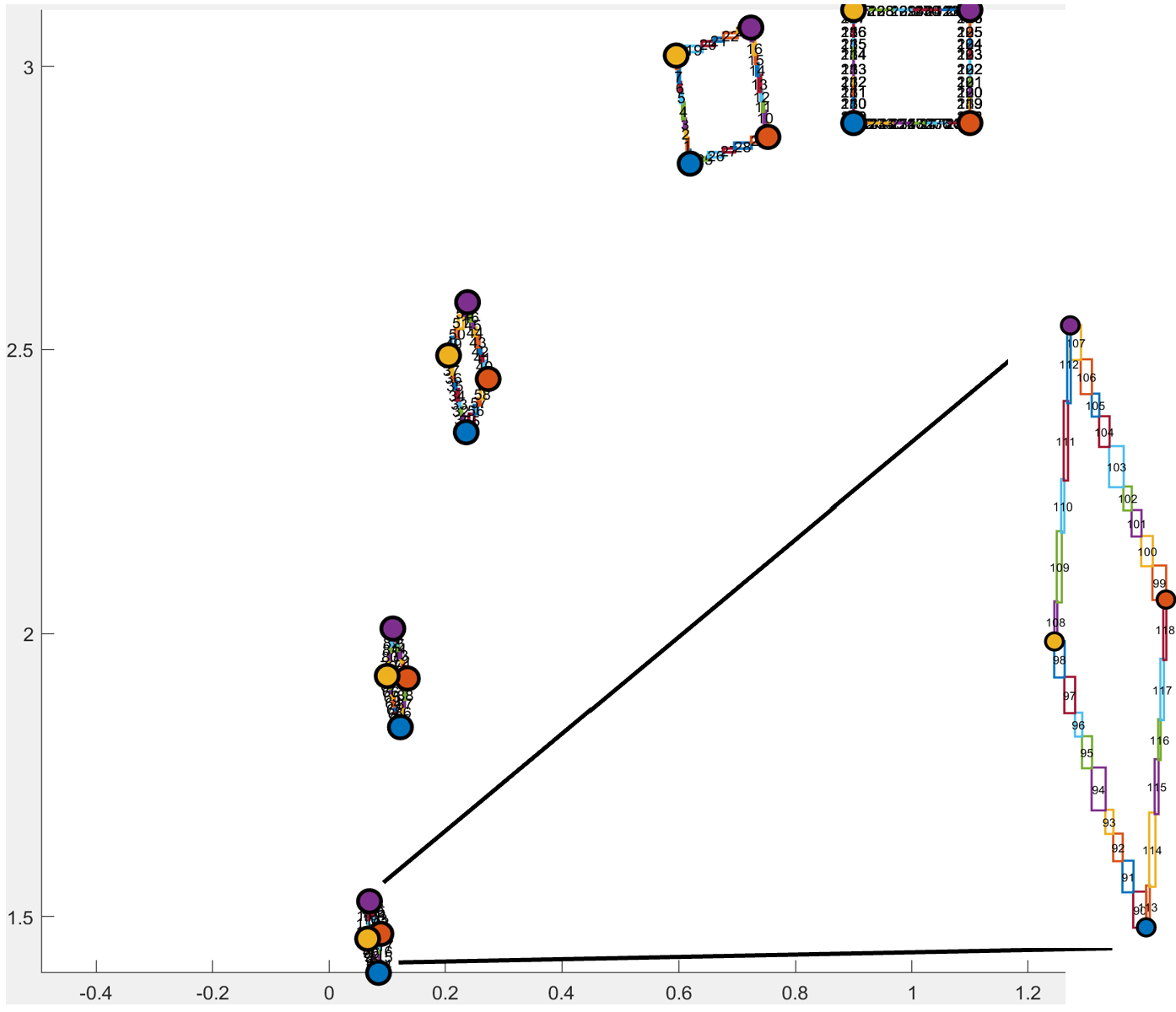}
			
			\vspace{4pt}
			{\footnotesize
				(a) Eg1: Results at $(T = 0, 0.1, 0.4, 0.7, 1.0)$.
				An enlarged view at $(T = 0.0, 1.0)$ is also shown.}
		\end{minipage}\hfill
		\begin{minipage}[t]{0.48\linewidth}
			\vspace{0pt}
			\centering
			\includegraphics[height=6.2cm]
			{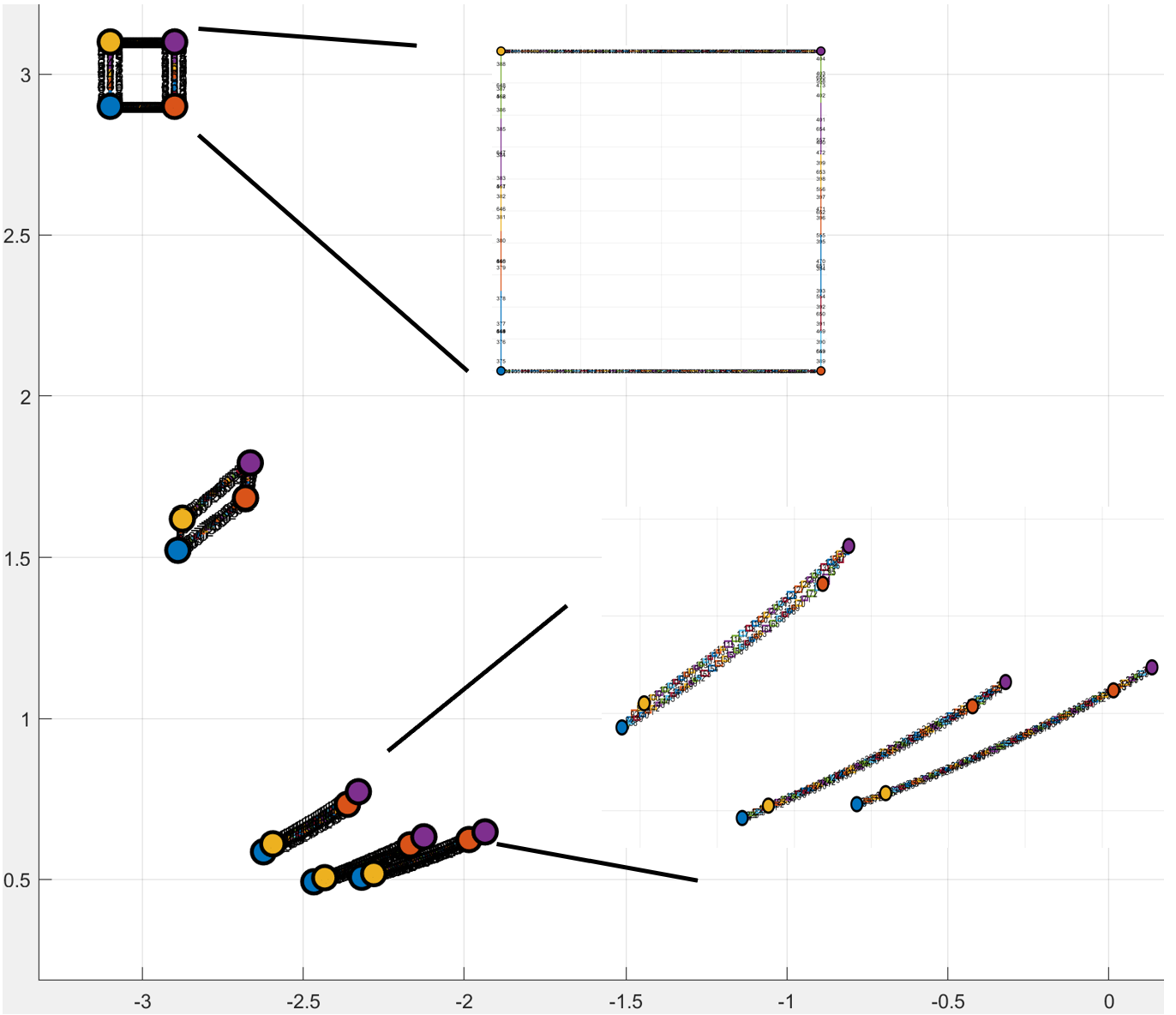}
			
			\vspace{4pt}
			{\footnotesize
				(b) Eg2: Results at $(T = 0, 0.1, 0.4, 0.7, 1.0)$.
				An enlarged view at $(T = 0.4, 0.7, 1.0)$ is also shown.}
		\end{minipage}
		
		\caption{Trajectories and end enclosures for Examples Eg1 and Eg2
			with $\veps=0.01$.}
		    \Description{Side-by-side trajectories and end enclosures for examples Eg1 and Eg2 at five times, with enlarged views of selected times.}
		\label{fig:eg1-eg2}
	\end{figure*}
	\begin{figure*}
		\centering
		\begin{minipage}[t]{0.48\linewidth}
			\centering
			\includegraphics[width=\linewidth, height=6cm]
			{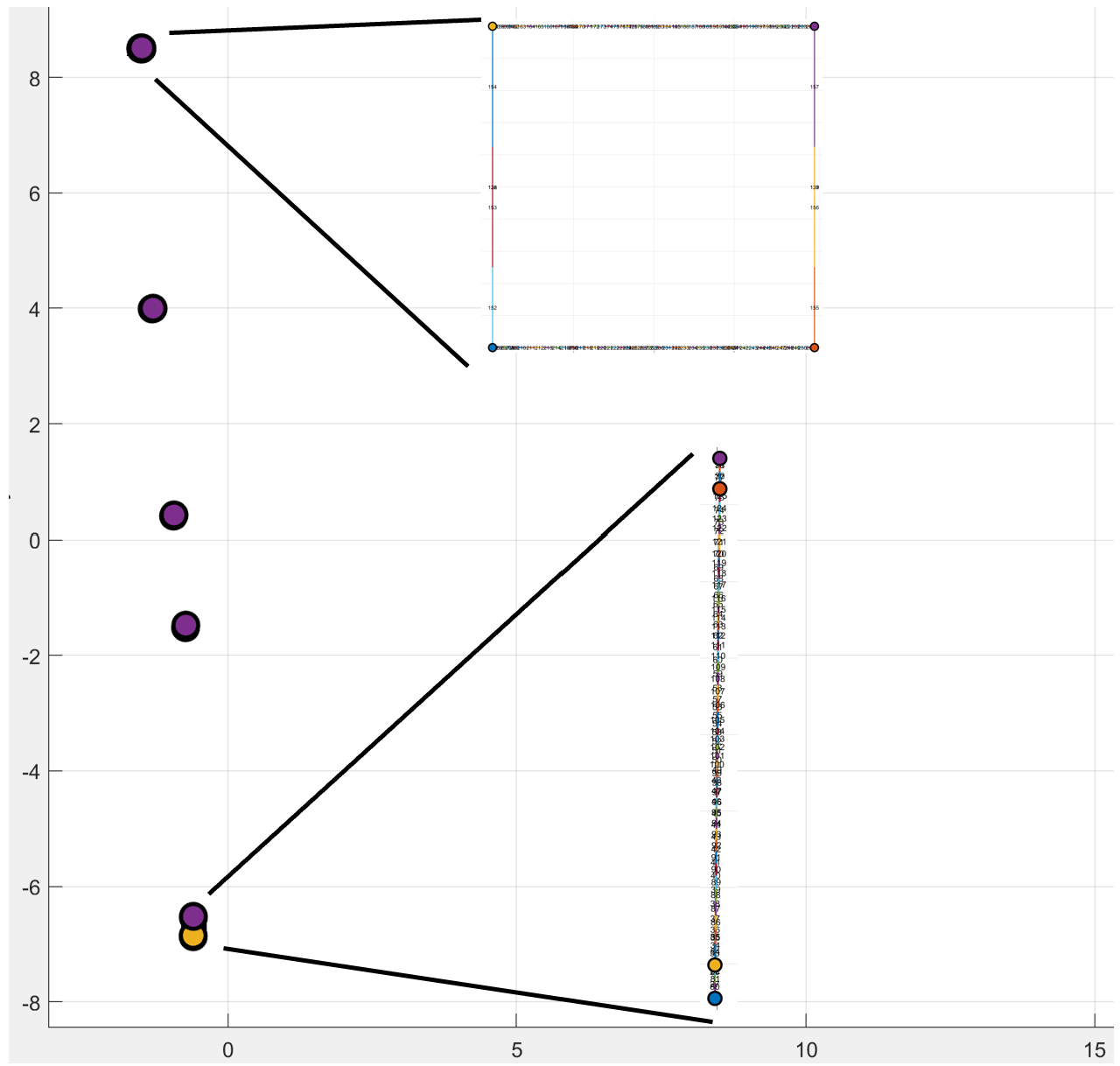}
			\vspace{4pt}
			
			{\footnotesize
				(c) Eg3: Results at $(T = 0, 0.1, 0.4, 0.7, 1.0)$.
				An enlarged view at $(T = 0.0, 1.0)$ is also shown.}
		\end{minipage}\hfill
		\begin{minipage}[t]{0.48\linewidth}
			\centering
			\includegraphics[width=\linewidth, height=6cm]
			{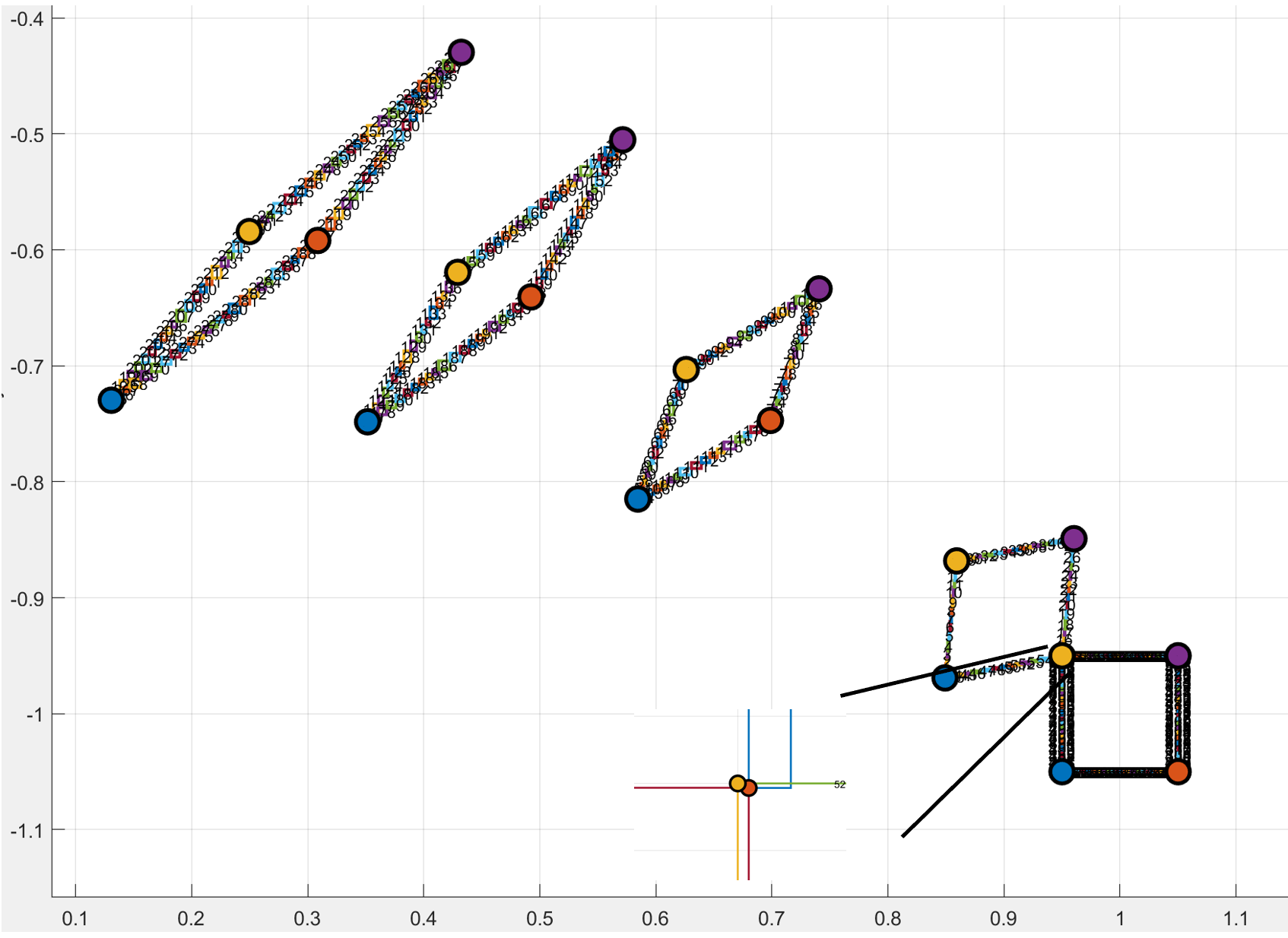}
			\vspace{4pt}
			
			{\footnotesize
				(d) Eg4: Results at $(T = 0, 0.1, 0.4, 0.7, 1.0)$.
				An enlarged view of the connected component at $(T = 0.0,
				0.1)$ is also shown.}
		\end{minipage}
		
		\caption{Trajectories and end enclosures for Example Eg3 and Eg4
			with $\veps=0.01$.}
		    \Description{Side-by-side trajectories and end enclosures for examples Eg3 and Eg4 at five times, with enlarged views of selected times.}
		\label{fig:eg3-eg4}
	\end{figure*}	
	\begin{figure*}
		\centering
		\begin{minipage}[t]{0.48\linewidth}
			\centering
			\includegraphics[width=\linewidth, height=6cm]
			{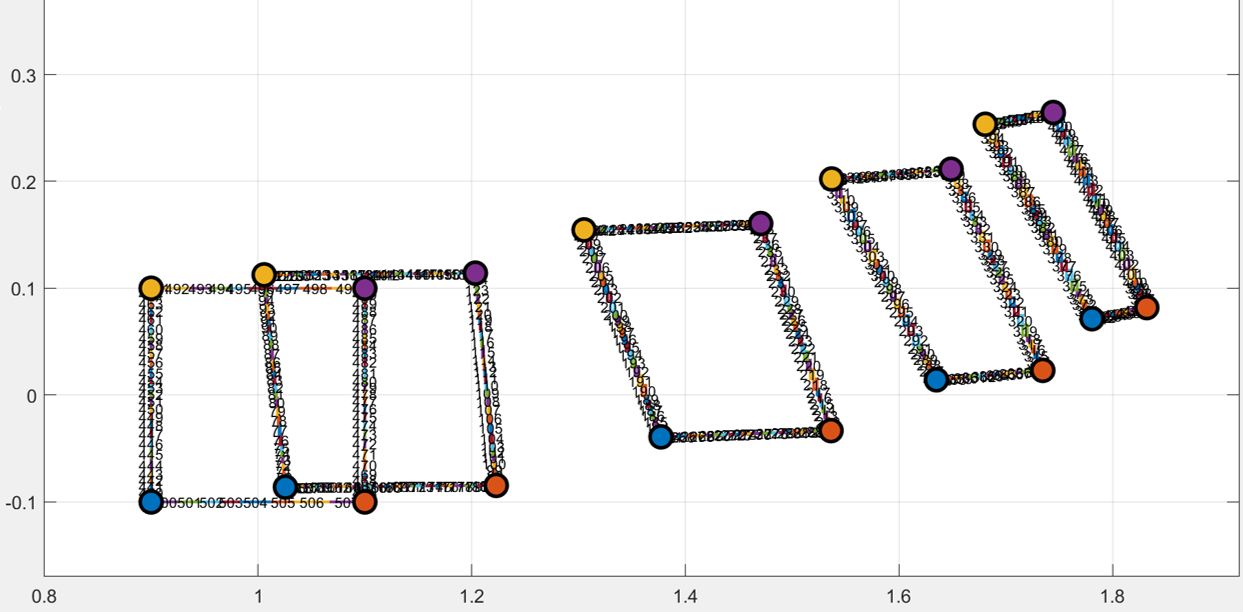}
			\vspace{4pt}
			
			{\footnotesize
				(c) Eg5: Results at $(T = 0, 0.1, 0.4, 0.7, 1.0)$.}
		\end{minipage}\hfill
		\begin{minipage}[t]{0.48\linewidth}
			\centering
			\includegraphics[width=\linewidth, height=6cm]
			{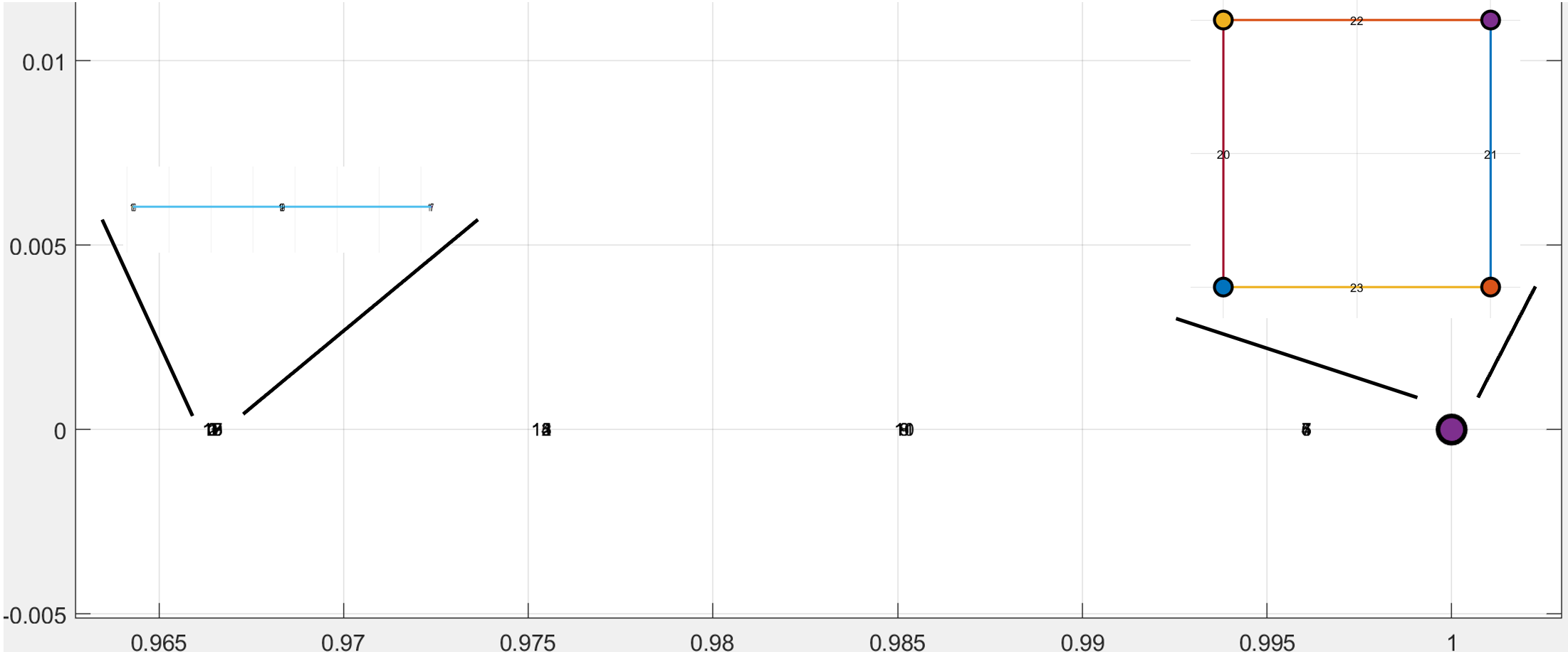}
			\vspace{4pt}
			
			{\footnotesize
				(d) Eg6: Results at $(T = 0, 0.1, 0.4, 0.7, 1.0)$.
				An enlarged view of the connected component at $(T = 0.0,
				1.0)$ is also shown.}
		\end{minipage}
		
		\caption{Trajectories and end enclosures for Example Eg5 and Eg6
			with $\veps=0.01$.}
		    \Description{Side-by-side trajectories and end enclosures for examples Eg5 and Eg6 at five times, including enlarged views.}
		\label{fig:eg5-eg6}
	\end{figure*}
	From these figures, we observe that, in contrast to Lohner’s method,
	our approach mitigates the wrapping effect by explicitly tracking and
	propagating the boundary of the reachable set.
	This boundary-oriented strategy prevents excessive over-approximation
	during enclosure propagation, which explains why our method is able to
	produce significantly tighter and more accurate end enclosures.

\issacArxiv[]{
	\ssect{Time Horizon Sensitivity Analysis}
		Table~\ref{tab:time} summarizes how the output 
	enclosure and computational effort depend on the time 
	horizon $T$. In general, increasing $T$ tends to 
	enlarge the computed enclosure and to increase the 
	number of initial boxes $\#(\ulB_0)$ and run time. 
	The effect can be mild for some problems (e.g., the 
	Quadratic example shows only modest changes in time 
	and $\#(\ulB_0)$), but can be drastic for others: Eg2 
	at $T=4.0$ requires 56 initial boxes and about 
	$20.36\,$s, whereas at $T\le 2.0$ it used at most 10 
	boxes and under $0.31\,$s. This indicates that long 
	time propagation may require significantly more 
	subdivision and thus higher computational cost; 
		\begin{table*}
			\centering
			\begin{tabular}{c|c|c|c|c}
				\hline
				\textbf{Case} & $T$ 
				& $Box(\olB_1)$ & $\#(\ulB_0)$ & \textbf{Time (s)} \\
				\hline
				Eg1 & 0.5
				& $[0.152,0.212]\times[2.170,2.391]$ & 8
				& 0.047 \\
				& 1.0
				& $[0.062,0.092]\times[1.395,1.532]$ & 8
				& 0.062 \\
				& 2.0
				& $[0.064,0.104]\times[0.553,0.601]$ & 8
				& 0.061 \\
				& 4.0
				& $[1.052,1.861]\times[0.136,0.240]$ & 12
				& 0.083 \\
				\hline
				Eg2 & 0.5
				& $[-2.586,-2.240]\times[0.464,0.748]$ & 8
				& 0.044 \\
				& 1.0
				& $[-2.3781-1.876]\times[0.365,0.781]$ & 8
				& 0.086 \\
				& 2.0
				& $[-1.783,-1.028]\times[0.607,1.312]$ & 10
				& 0.152 \\
				& 4.0  
				& $[1.351,2.272]\times[-0.343,2.470]$ & 56
				& 2.361 \\
				\hline
				Eg3 & 0.1
				& $[-1.311,-1.296]\times[3.986,4.000]$ & 8
				& 0.050 \\
				& 0.4
				& $[-0.941,-0.933]\times[0.398,0.419]$ & 8
				& 0.065\\
				& 0.7
				& $[-0.734,-0.729]\times[-1.519,-1.474]$ & 8
				& 0.073 \\
				& 1.0
				& $[-0.601,-0.598]\times[-6.877,-6.508]$ & 8
				& 0.082\\
				\hline
				Quadratic & 0.5
				& $[0.503,0.680]\times[-0.789,-0.581]$ & 8
				& 0.018\\
				& 1.0
				& $[0.125,0.434]\times[-0.743,-0.424]$ & 8
				& 0.024\\
				& 2.0
				& $[-0.620,0.069]\times[-0.707,-0.314]$ & 8
				& 0.061\\         
				& 4.0
				& $[-0.888,-0.411]\times[-0.318,0.828]$ & 11
				& 0.059\\
				\hline
			\end{tabular}
			\caption{Performance variation of the  algorithm under different time target  $T$.
				$\veps=1.0$.}
			\label{tab:time}
		\end{table*}

}

\ssect{Tolerance Parameter Sensitivity Analysis}	
Table \ref{tab:veps} analyzes the impact of varying the
tolerance parameter $\veps$.  As seen in the last two
columns of the table, as $\veps$ decreases, the width
of the end-enclosure decreases (as expected), but the
running time increases (as expected).

\begin{table*}
	\centering
	\begin{tabular}{c|c|c|c|c}
		\hline
		\textbf{Case} & $\veps$ 
		& End-Enclosure $B_1$ & $\wmax(B_1)$ & \textbf{Time (s)} \\
		\hline
		Eg1 & 1.0
		& $[0.062,0.092]\times [1.395,1.532]$ & 0.137
		& 0.042 \\
		& 0.5
		& $[0.062,0.092]\times [1.395,1.532]$ & 0.137
		& 0.042 \\
		& 0.1
		& $[0.062,0.092]\times [1.395,1.532]$ & 0.137
		& 0.042 \\
		& 0.01
		& $[0.066,0.089]\times[1.400,1.527]$ & 0.126
		& 0.874 \\
		\hline
		Eg2 & 1.0
		& $[-2.381,-1.876]\times[0.365,0.781]$ & 0.505
		& 0.083 \\
		& 0.5
		& $[-2.381,-1.876]\times[0.365,0.781]$ & 0.505
		& 0.080 \\
		& 0.1
		& $[-2.321,-1.933]\times[0.503,0.651]$ & 0.387
		& 0.270 \\
		& 0.01
		& $[-2.318,-1.935]\times[0.506,0.647]$ & 0.383
		& 3.791 \\
		\hline
		Eg3 & 1.0
		& $[-0.601,-0.598]\times[-6.877,-6.508]$ & 0.369
		& 0.090 \\
		& 0.5
		& $[-0.601,-0.598]\times[-6.877,-6.508]$ & 0.369
		& 0.098 \\
		& 0.1
		& $[-0.601,-0.598]\times[-6.872,-6.519]$ & 0.353
		& 0.293 \\
		& 0.01
		& $[-0.601,-0.598]\times[-6.872,-6.519]$ & 0.352
		& 6.003 \\
		\hline
		Eg4 & 1.0 & $[0.125,0.434]\times[-0.743,-0.424]$&0.319&
		0.021\\
		&   0.5 & $[0.125,0.434]\times[-0.743,-0.424]$&0.319& 0.021\\
		& 0.1 & $[0.129,0.432]\times[-0.733,-0.428]$&
		0.304 & 0.046\\
		& 0.01& $[0.130,0.432]\times[-0.729,-0.429]$&
		0.301& 0.885\\
		\hline
		Eg5 & 1.0 & $[1.649,1.864]\times[0.687,0.268]$
		& 0.215 & 0.078\\
		& 0.5 & $[1.646,1.866]\times[0.068,0.268]$ &
		0.220 & 0.070\\
		& 0.1 & $[1.678,1.835]\times[0.070,0.265]$&
		0.195 & 0.169\\
		& 0.01 & $[1.680,1.831]\times [0.071,0.264]$ &
		0.193& 2.400\\
		\hline
		Eg7 & 1.0 &
		$[-6.976,-6.913]\times[2.986,3.007]\times[35.104,35.184]$
		& 0.079 & 0.647\\
		& 0.5 &
		$[-6.976,-6.913]\times[2.986,3.007]\times[35.104,35.184]$ &
		0.079 & 0.647\\
		& 0.1 &
		$[-6.976,-6.913]\times[2.986,3.007]\times[35.104,35.184]$ &
		0.079 & 0.647\\
		& 0.05&
		$[-6.976,-6.914]\times[2.987,3.006]\times[35.105,35.183]$ &
		0.078 & 6.083\\
		\hline
		Eg8 & 1.0 &
		$[-1.880,-1.588]\times[1.702,2.039]\times[0.031,0.035]$
		& 0.337 & 0.162\\
		& 0.5 &
		$[-1.880,-1.588]\times[1.702,2.039]\times[0.031,0.035]$
		& 0.337 & 0.162\\
		& 0.1 &
		$[-1.876,-1.591]\times[1.705,2.036]\times[0.032,0.035]$
		& 0.331 & 4.574\\
		& 0.05 &
		$[-1.875,-1.592]\times[1.706,2.036]\times[0.032,0.035]$
		& 0.329 & 36.810\\
		\hline
	\end{tabular}
	\caption{Performance variation of the algorithm under different
		tolerance parameters $\veps$.}
	\label{tab:veps}
\end{table*}

\ignore{
	\textbf{Experimental Conclusion Summary:} Table \ref{tab:veps} shows
	that decreasing the tolerance parameter $\veps$ generally leads to
	contraction of interval widths (e.g., $\wmax$ decreases from 0.20 to
	0.12 in Eg1), but at the cost of increased computation time. Notably,
	when $\veps$ falls below a certain threshold (e.g., $\veps \leq 0.1$
	in Eg2 and Eg3), the interval width no longer changes significantly,
	indicating that the algorithm has reached its precision limit. This
	phenomenon provides guidance for selecting appropriate $\veps$ values
	in practical applications: excessively small $\veps$ may
	unnecessarily increase computational costs without improving
	accuracy.
}%

\section{End Covers, Convexification, and the Wrapping Effect}
\label{app:classical-examples}

This appendix uses two classical examples to clarify how the
EndCover formulation differs from the usual propagation of a single
global enclosure.
The examples expose two distinct sources of overestimation.
Eijgenraam's example illustrates a \emph{spatial convexification}
obstruction: even an exact computation of the flow cannot represent
a curved end set tightly by one convex enclosure.
Moore's rotation example illustrates a \emph{temporal wrapping}
obstruction: repeated validated propagation and re-boxing may enlarge
an enclosure even when the exact flow preserves its size.

\subsection{Eijgenraam's Example:
A Limitation of One Convex Enclosure}
\label{app:eijgenraam}

Consider Eijgenraam's classical example \cite[Example 6.4.2, pg.~127]{eijgenraam:ivp:81} (also \cite[p.~77-78]{nedialkov:thesis:99}): 
\begin{equation}
	\label{eq:app-eijgenraam-ode}
	x'=0,
	\qquad
	y'=x^2,
	\qquad
	B_0=[-\lambda,\lambda]\times\{0\}.
\end{equation}
For an initial point $(\xi,0)$, the exact solution is
\[
	x(t)=\xi,
	\qquad
	y(t)=t\xi^2.
\]
Thus the exact end set at time $H$ is
\begin{equation}
	\label{eq:app-eijgenraam-end}
	\Gamma_H
	:=
	\End_{\bff}(B_0,H)
	=
	\{(\xi,H\xi^2):|\xi|\leq\lambda\},
\end{equation}
which is a parabolic arc.

The important point is that the obstruction remains even if the ODE
flow is computed exactly.
Let $K$ be any convex set containing $\Gamma_H$.
Since $K$ contains
\[
	(-\lambda,H\lambda^2)
	\qquad\hbox{and}\qquad
	(\lambda,H\lambda^2),
\]
convexity forces it to contain their midpoint
\begin{equation}
	z_*=(0,H\lambda^2).
\end{equation}
However, $z_*\notin\Gamma_H$.

More quantitatively,
\begin{align}
	d_\infty(z_*,\Gamma_H)
	&=
	\min_{0\leq u\leq\lambda}
	\max\{u,H(\lambda^2-u^2)\}
	\nonumber\\
	&=
	\frac{\sqrt{1+4H^2\lambda^2}-1}{2H}
	\nonumber\\
	&=
	H\lambda^2+O(\lambda^4).
	\label{eq:app-eijgenraam-lower}
\end{align}
Therefore every single convex enclosure has a nonzero unavoidable
excess.
For $H=\lambda=1$,
\[
	d_\infty(z_*,\Gamma_1)
	=
	\frac{\sqrt{5}-1}{2}
	\approx 0.618.
\]

For an axis-aligned box, the situation is even more transparent.
The smallest such box containing the exact end set is
\[
	[-\lambda,\lambda]\times[0,H\lambda^2].
\]
Thus the large region between this box and the parabola is not caused
by truncation, rounding, or an inaccurate validated solver.
It results solely from forcing a nonconvex end set into one convex
representation.

\begin{figure*}[t]
	\centering
	\includegraphics[width=0.98\textwidth]
		{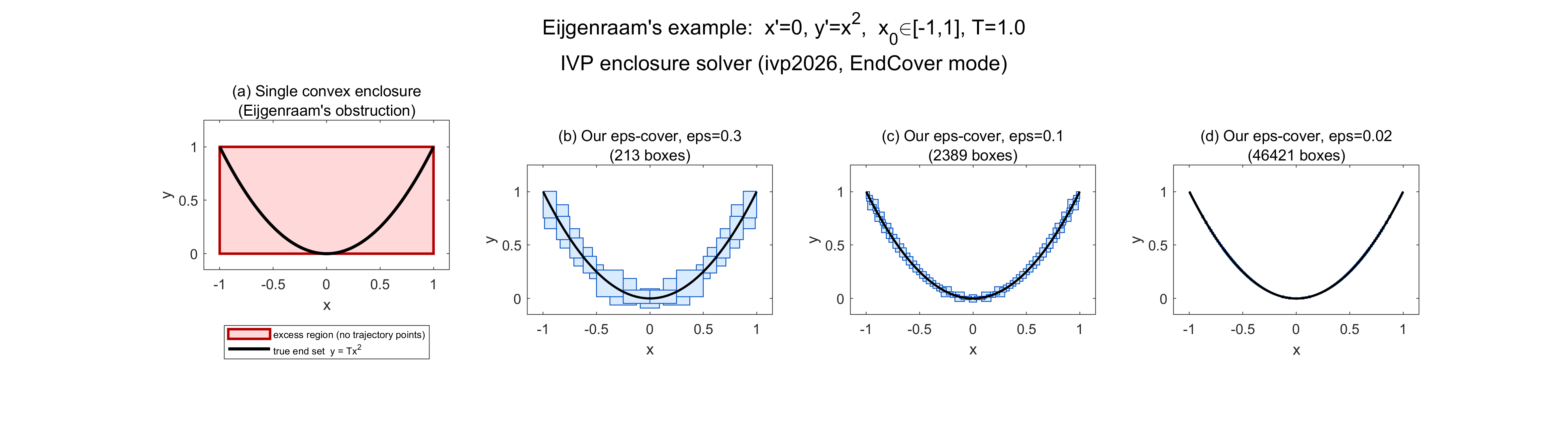}
	\caption{
	Eijgenraam's example
	$x'=0$, $y'=x^2$,
	$x_0\in[-1,1]$, and $H=1$.
	(a) The black curve is the exact end set
	$y=x^2$.  The red rectangle is its smallest axis-aligned
	box, and the shaded region consists of points that are not in
	the true end set.
	(b)--(d) EndCover outputs for
	$\varepsilon=0.3$, $0.1$, and $0.02$,
	containing respectively
	$213$, $2389$, and $46421$ boxes.
	}
	\Description{The exact parabolic end set and its axis-aligned box are compared with three EndCover approximations; smaller tolerances use more boxes and follow the curve more closely.}
	\label{fig:app-eijgenraam}
\end{figure*}

EndCover avoids this geometric obstruction because
$\bigcup_{B\in C}B$ need not be convex.
For the experiment in Figure~\ref{fig:app-eijgenraam}, we use
\[
	\lambda=1,
	\qquad
	H=1.
\]
We run EndCover for three values of $\varepsilon$:

\begin{center}
\begin{tabular}{c|ccc}
	$\varepsilon$
	& $0.3$ & $0.1$ & $0.02$
	\\ \hline
	Number of boxes $|C|$
	& $213$ & $2389$ & $46421$
\end{tabular}
\end{center}

As $\varepsilon$ decreases, the union of output boxes follows the
parabolic end set increasingly closely.

\subsection{Two methods to control the
Moore's wrapping effect.
}
\label{app:moore}

We next consider Moore's classical rotation example
\begin{equation}
	\label{eq:app-moore-ode}
	x'=y,
	\qquad
	y'=-x,
	\qquad
	B_0=[0.95,1.05]\times[-0.05,0.05].
\end{equation}
The experiments below use
\begin{equation}
	\varepsilon=0.02.
\end{equation}

Writing
\[
	R(\theta)
	=
	\begin{pmatrix}
		\cos\theta & -\sin\theta\\
		\sin\theta &  \cos\theta
	\end{pmatrix},
\]
the exact end set is
\begin{equation}
	\label{eq:app-moore-exact}
	\End_{\bff}(B_0,H)
	=
	R(-H)B_0.
\end{equation}
Hence the exact solution set is simply a rigidly rotating square:
its diameter and area remain constant for all $H$.

This example exposes a different phenomenon from
Subsection~\ref{app:eijgenraam}.
Suppose that after a rotation by $\theta$ a square is replaced by
its axis-aligned box hull.
Its width is multiplied by
\begin{equation}
	|\cos\theta|+|\sin\theta|.
\end{equation}
For $\theta=\pi/4$, this factor is $\sqrt{2}$.
Thus, if a box of width $0.1$ is propagated through eight successive
$\pi/4$ rotations and re-boxed after every step, its width can grow to
\[
	0.1(\sqrt{2})^8=1.6
\]
after one complete revolution, although the exact rotated square
still has width $0.1$.
This is the classical wrapping effect.

To distinguish this internal propagation effect from the EndCover
formulation itself, we perform two EndCover experiments on exactly
the same problem and with the same $\varepsilon=0.02$.
The only important difference is the StepB propagation mechanism.

\subsubsection{EndCover with the Pure Direct Method}
\label{app:moore-direct}

In the first experiment, EndCover is stepped using the
\emph{ Direct method}.

We evaluate the cover at
\begin{equation}
	H
	\in
	\left\{
	0,\frac{\pi}{4},\frac{\pi}{2},\frac{3\pi}{4},
	\pi,\frac{5\pi}{4},\frac{3\pi}{2},\frac{7\pi}{4},2\pi
	\right\}.
\end{equation}

The experimentally observed numbers of output boxes are
\begin{equation}
	\label{eq:app-moore-direct-counts}
	85,\,
	85,\,
	341,\,
	341,\,
	341,\,
	1365,\,
	1365,\,
	5461,\,
	5461.
\end{equation}

Thus, although the true set never grows, progressively more boxes
are required to maintain the prescribed $\varepsilon$-cover as
elapsed time increases.

This is an important distinction.
EndCover still succeeds in keeping the \emph{output cover} localized
near the exact rotating square.
However, because the Direct propagation loses orientation
information, EndCover must compensate for the accumulated wrapping
by increasingly fine spatial subdivision.

\begin{figure}[t]
	\centering
	\includegraphics[width=0.42\textwidth]
		{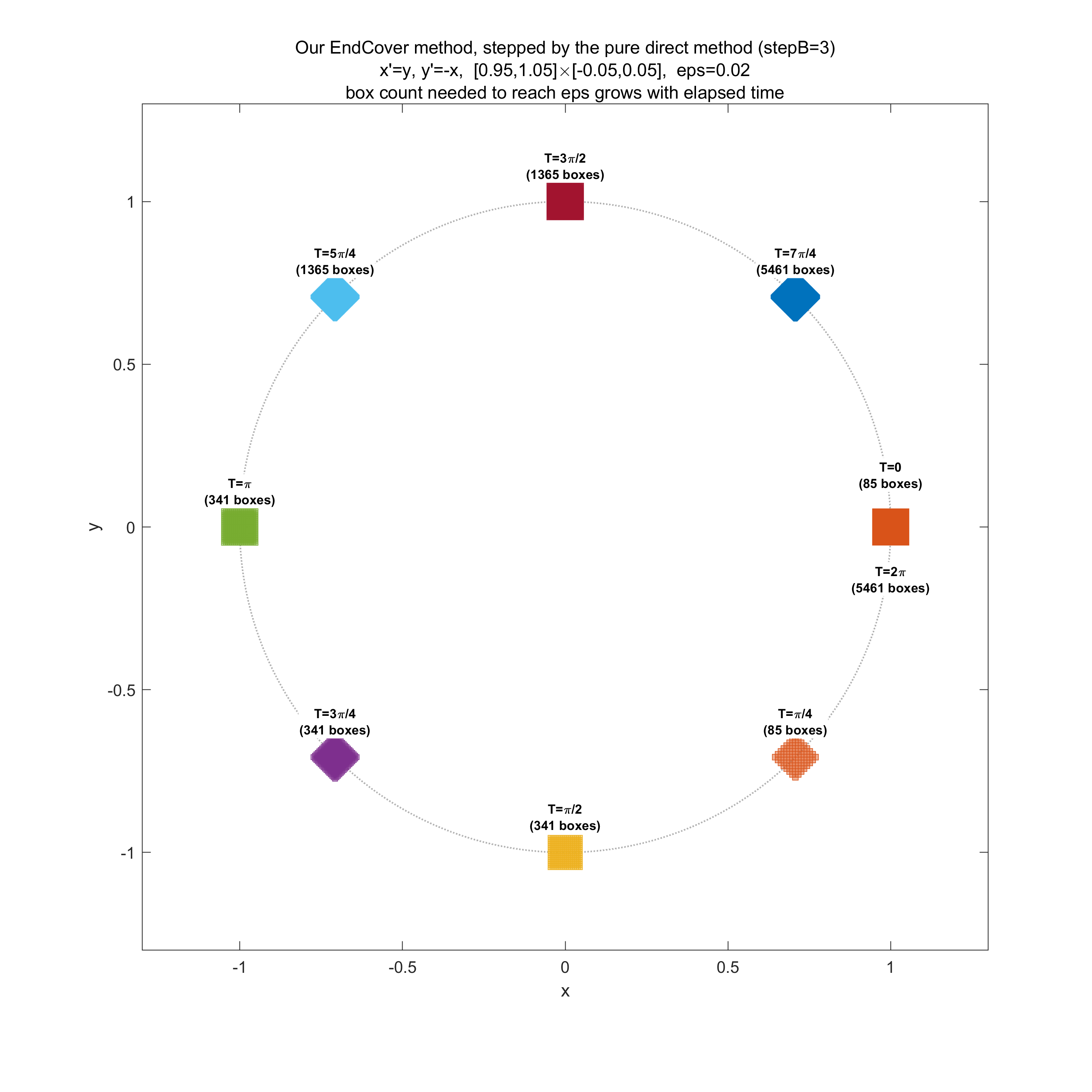}
	\caption{
	Moore's rotation example with
	$B_0=[0.95,1.05]\times[-0.05,0.05]$ and
	$\varepsilon=0.02$, using EndCover with the pure Direct
	StepB method.
	The exact orbit has no growth, but the number of output boxes
	required to maintain the $\varepsilon$-cover increases with
	elapsed time:
	$85,85,341,341,341,1365,1365,5461,5461$
	at
	$H=0,\pi/4,\ldots,2\pi$.
	}
	\Description{Box counts for the Direct-method cover of a rotating square increase over one revolution, although the exact square keeps constant size.}
\end{figure}
\label{fig:app-moore-direct-overall}

Figure~\ref{fig:app-moore-direct-panels} shows the internal geometry
of these covers.
The black outline denotes the exact rotated square and the blue
rectangles are the computed EndCover boxes.

\begin{figure}
	\centering
	\includegraphics[width=0.46\textwidth]
		{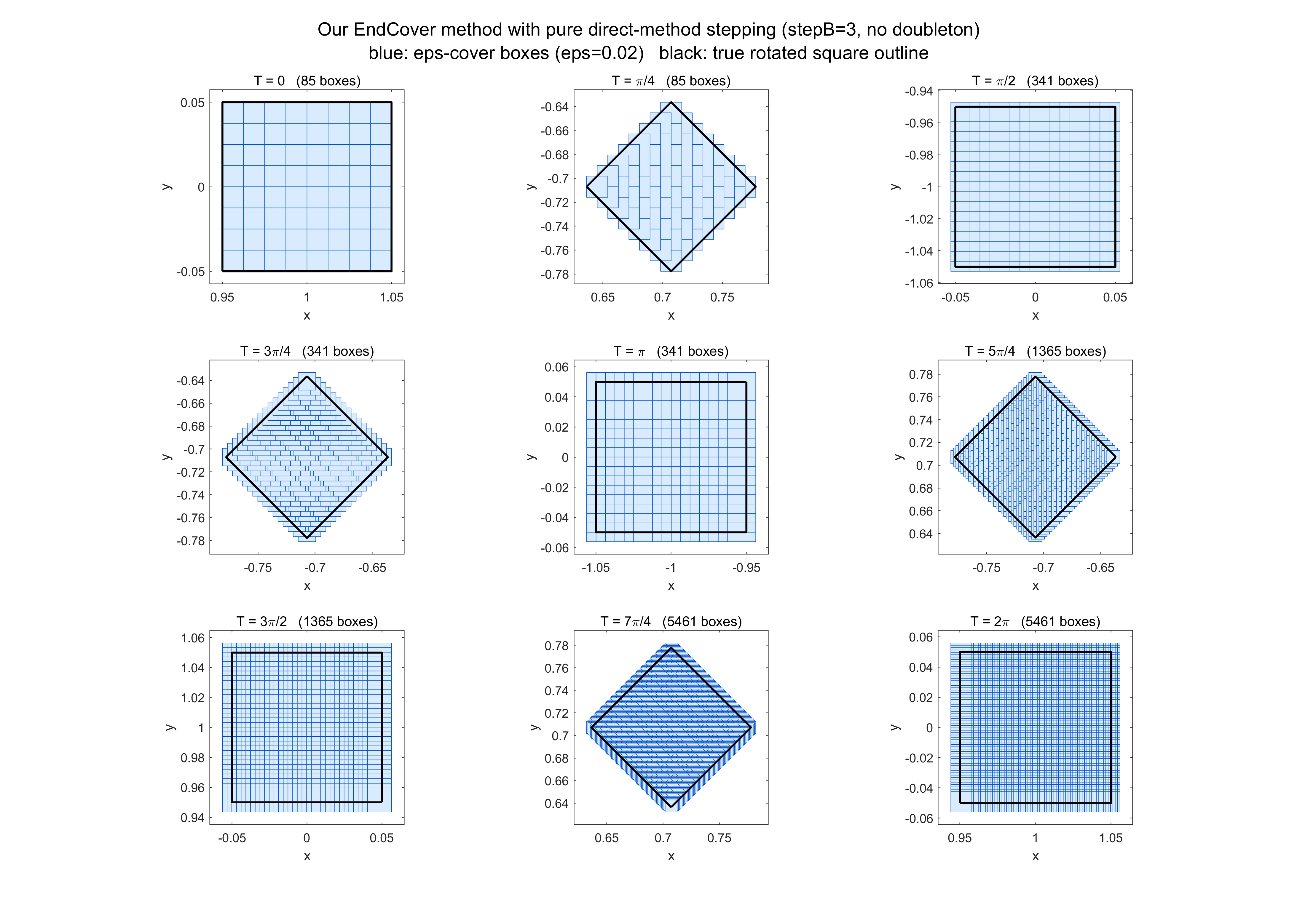}
	\caption{
	Detailed EndCover outputs for the pure Direct-method experiment
	in Figure~\ref{fig:app-moore-direct-overall}.
	The black outline is the exact rotated square and the blue
	rectangles are the output $\varepsilon$-cover boxes.
	The number of boxes is
	$85$ at $H=0,\pi/4$;
	$341$ at $H=\pi/2,3\pi/4,\pi$;
	$1365$ at $H=5\pi/4,3\pi/2$; and
	$5461$ at $H=7\pi/4,2\pi$.
	The increasing subdivision compensates for wrapping accumulated
	by the pure box-based propagation.
	}
	\Description{Snapshots of the Direct-method cover show increasingly many blue boxes around the exact rotated square as wrapping accumulates.}
\end{figure}
\label{fig:app-moore-direct-panels}

\subsubsection{EndCover with Lohner-Type Propagation}
\label{app:moore-lohner}

We next run the same EndCover problem while using a Lohner-type
propagation that retains an internal moving representation of the
solution set.
The EndCover output is still a collection of axis-aligned boxes;
what changes is the representation used to propagate the set from
one step to the next.
Table~\ref{tab:app-moore-counts} compare the box count from the two experiments.

\begin{table}
	\centering
	\caption{
	Number of EndCover output boxes for Moore's rotation example
	with $\varepsilon=0.02$.
    Comparing Direct method with Lohner's method.
	}
	\label{tab:app-moore-counts}
	\setlength{\tabcolsep}{4.5pt}
	\begin{tabular}{c|rrrrrrrrr}
		\hline
		$H$
		& $0$
		& $\pi/4$
		& $\pi/2$
		& $3\pi/4$
		& $\pi$
		& $5\pi/4$
		& $3\pi/2$
		& $7\pi/4$
		& $2\pi$
		\\ \hline

		Direct
		& $85$
		& $85$
		& $341$
		& $341$
		& $341$
		& $1365$
		& $1365$
		& $5461$
		& $5461$
		\\

		Lohner
		& $85$
		& $85$
		& $85$
		& $85$
		& $85$
		& $85$
		& $85$
		& $85$
		& $85$
		\\
		\hline
	\end{tabular}
\end{table}

The following distinction is important:
The EndCover algorithm determines the global spatial-cover logic and
the prescribed $\varepsilon$ guarantee.
The Lohner representation controls wrapping during the local
validated propagation.
They therefore operate at different levels of the algorithm.

\begin{figure}
	\centering
	\includegraphics[width=0.42\textwidth]
		{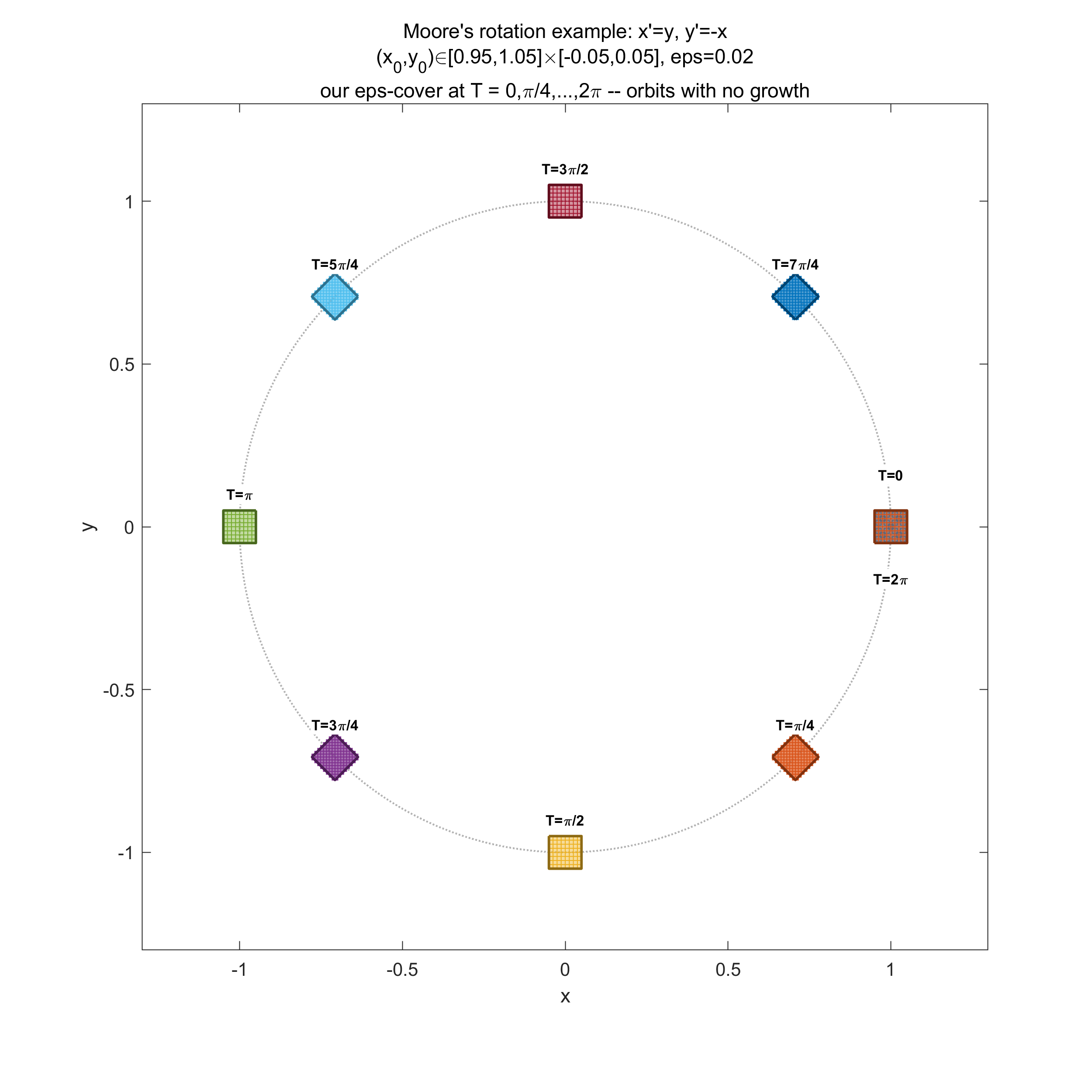}
	\caption{
	Moore's rotation example with
	$B_0=[0.95,1.05]\times[-0.05,0.05]$ and
	$\varepsilon=0.02$, using EndCover with Lohner-type
	validated propagation.
	The cover is shown at
	$H=0,\pi/4,\ldots,2\pi$.
	The output contains $85$ boxes at every snapshot;
	there is no growth in cover size over one complete revolution.
	}
	\Description{Snapshots of the Lohner-type cover show 85 blue boxes around the exact rotated square at every angle over one revolution.}
	\label{fig:app-moore-lohner-overall}
\end{figure}

\begin{figure}
	\centering
	\includegraphics[width=0.46\textwidth]
		{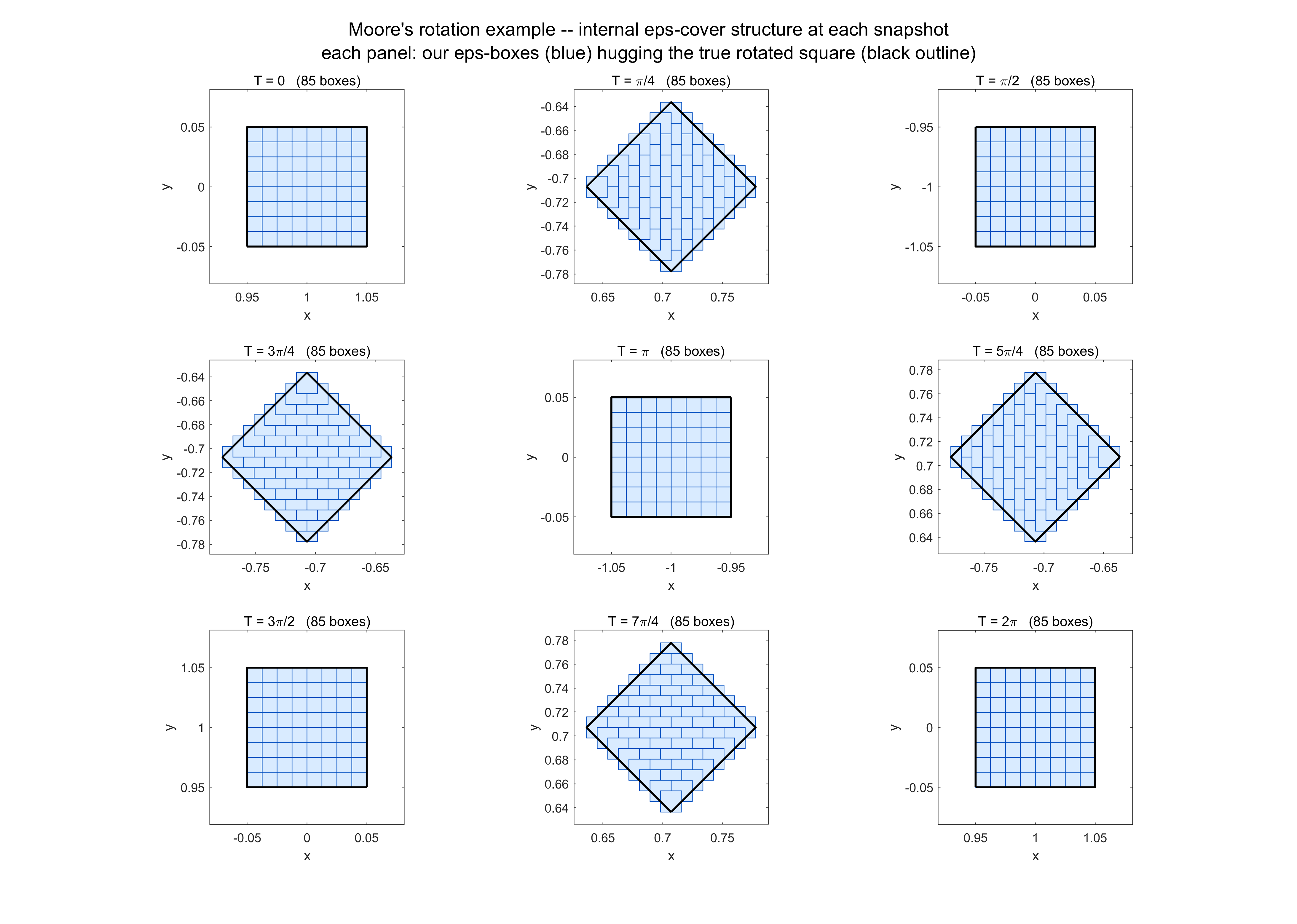}
	\caption{
	Detailed structure of the Lohner-enhanced EndCover outputs.
	At each of
	$H=0,\pi/4,\pi/2,3\pi/4,\pi,5\pi/4,3\pi/2,7\pi/4,2\pi$,
	the cover contains exactly $85$ boxes.
	The black outline is the exact rotated square and the blue
	rectangles are the EndCover boxes.
	The orientation of the solution set is retained internally during
	propagation rather than discarded after every step.
	}
	\Description{Detailed Lohner-type cover snapshots show the exact rotated square and 85 blue boxes at each of nine angles.}
	\label{fig:app-moore-lohner-panels}
\end{figure}
